\documentclass{aa}
\usepackage{graphicx}
\usepackage{amsmath}	
\usepackage{amssymb}	
\usepackage{multicol}        
\usepackage{bm}		
\usepackage{pdflscape}	
\usepackage{ulem} 
\usepackage{xcolor}
\usepackage{esint}
\usepackage{sidecap}
\usepackage[flushleft]{threeparttable}  
\newcommand{\vsini}{$v$~sin$i$}
\newcommand{\chit}{$\chi^{2}$}
\newcommand{\chir}{$\chi^{2}_r$}
\newcommand{\halpha}{H$\alpha$}
\newcommand{\teff}{$T_\mathrm{eff}$}

\newcommand{\degrr}{$^{\circ}$}

\newcommand{\logg}{$log~g$}
\newcommand{\ap}{V530~Per}
\newcommand{\sn}{S/N}
\newcommand{\StI}{Stokes \textit{I}}
\newcommand{\StV}{Stokes \textit{V}}

\def\kms{\hbox{km\,s$^{-1}$}}

\def\gt{\textgreater}

\begin{document}

\title{
Short-term variations of surface magnetism and prominences of the young Sun-like star V530 Per}
\titlerunning{Short-term variations of V530 Per}
\author{T.-Q. Cang \inst{1} 
\and P. Petit \inst{1} 
\and J.-F. Donati \inst{1}
\and C.P. Folsom \inst{2}
}

\institute
{Institut de Recherche en Astrophysique et Plan\'etologie, Universit\'e de Toulouse, CNRS, CNES, 14 avenue Edouard Belin, 31400 Toulouse, France \\
\email{Tianqi.Cang@irap.omp.eu}
\and
Tartu Observatory, University of Tartu, Observatooriumi 1, T\~{o}ravere, 61602 Tartumaa, Estonia
}

\abstract
{}
{We investigate magnetic tracers in the photosphere and the chromosphere of the ultra-rapid rotator ($P\sim0.32d$) V530 Per, a cool member of the open cluster $\alpha$ Persei, to characterize the short-term variability of the magnetic activity and large-scale magnetic field of this prototypical young, rapidly rotating solar-like star. } 
{With time-resolved spectropolarimetric observations spread over four close-by nights, we reconstructed the brightness distribution and large-scale magnetic field geometry of V530 Per through Zeeman-Doppler imaging. Simultaneously, we estimated the short-term variability of the surface through latitudinal differential rotation. Using the same data set, we also mapped the spatial distribution of prominences through tomography of \halpha\ emission.}
{As in our previous study, a large dark spot occupies the polar region of V530 Per with smaller, dark, and bright spots at lower latitudes. The large-scale magnetic field is dominated by a toroidal, mostly axisymmetric component. The maximal radial field strength is equal to $\sim1$~kG. The surface differential rotation is consistent with a smooth Sun-like shear d$\Omega = 0.053 \pm 0.004$ rad.d$^{-1}$, close to the solar shear level. The prominence pattern displays a stable component that is confined close to the corotation radius. We also observe rapidly evolving \halpha\ emitting structures, over timescales ranging from minutes to days. The fast \halpha\ evolution was not linked to any detected photospheric changes in the spot or magnetic coverage.}
{}

\maketitle

\section{Introduction}
\label{sec:intro}

Stellar activity is an important aspect of the early evolution of solar-type stars. A large fraction of young Suns rotate rapidly \citep{2015A&A...577A..98G}, and the sustained rotation is the root of  strong activity, as empirically illustrated by the well-known  relationship between X-ray emission and the rotation rate (e.g., \citealt{2011ApJ...743...48W}). For stars at the higher end of the rotation rate distribution, however, the activity stops increasing with the rotation rate (e.g., \citealt{2011ApJ...743...48W, 2014MNRAS.441.2361V, 2019ApJ...876..118S}). This phenomenon is believed to be linked to a so-called saturation of the stellar dynamo. There were several theoretical attempts to model this specific dynamo state \citep{2015RAA....15.1801K,2017EPJWC.16002010A, 2019ApJ...880....6G}, but currently we still lack a detailed knowledge of dynamo saturation and few Sun-like dwarfs with a saturated large-scale field strength have benefited from a detailed investigation of their magnetic activity (LO Peg, \citealt{2016MNRAS.457..580F}; AB Dor, \citealt{2003MNRAS.345.1145D}; and BD-072388,  \citealt{2018MNRAS.474.4956F}).

One major physical ingredient responsible for the amplification of magnetic fields through global dynamos in cool stars is the differentially rotating stellar envelope. Zeeman-Doppler imaging (ZDI) of stellar surfaces  \citep{1989A&A...225..456S,2006MNRAS.370..629D,2018MNRAS.474.4956F} offers an efficient way to study the spot patterns and the large-scale magnetic field geometries of active stars, as well as the progressive modification of the surface tracers under the influence of differential rotation. The solar surface differential rotation measured through different photospheric tracers displays relatively consistent and stable shear levels \citep{2000SoPh..191...47B}. ZDI investigations of young, active stars reveal a more diverse situation, where temporal changes of differential rotation were observed, and shear values were reported to depend on the adopted surface tracer (brightness or magnetic field, e.g., \citealt{2003MNRAS.345.1187D, 2019MNRAS.489.5556Y}). 

Above the photosphere, magnetic fields create local accumulations of gas at chromospheric temperatures supported by magnetic loops and extending into the stellar corona. Stellar prominences can be used as a proxy to estimate the loss of mass and angular momentum through the stellar wind (e.g., \citealt{2019MNRAS.485.1448V, 2019MNRAS.482.2853J,2021LRSP...18....3V}), which is of prime importance in our understanding of the early evolution of cool stars. Although the knowledge we get from the Sun can provide rich information to understand the physics of prominences in a stellar context, available observations of prominence systems in other stars remain scarce as of today, so that the precise loss of mass and angular momentum through ejected prominence material is still uncertain. For rapidly rotating objects, the centrifugal force becomes involved in the equilibrium and dynamics of the so-called slingshot prominences (e.g.,  \citealt{1989MNRAS.236...57C,1989MNRAS.238..657C, 1999MNRAS.302..437D}). Slingshot prominences were witnessed in corotation with a few, young solar-like objects (e.g., \citealt{2000MNRAS.316..699D,2006MNRAS.373.1308D,2020A&A...643A..39C,2021MNRAS.504.1969Z}), and were shown to evolve over time scales as short as a few days. 

V530 Per (also named AP 149) is a member of the young open cluster $\alpha$ Persei \citep{1992AJ....103..488P}, with an age of around $60$~Myr \citep{2018A&A...615A..12Y}. It is a close analogue to the young Sun, with a mass of $1.00\pm0.05$~$M_{\sun}$, a radius of $1.06 \pm 0.11~R_{\sun}$, and an effective temperature of $5281\pm96$~K (\citealt{2020A&A...643A..39C}, C20 hereafter). As an ultra-rapid rotator with a rotational period P~$\sim0.32$d (Rossby number of $\sim0.013$), it lies in the saturated regime of the dynamo action, and may even reach super-saturation. The spectropolarimetric study of C20, based on data obtained in 2006, highlighted the surface distribution of brightness structures, as well as a complex magnetic field geometry. The rotationally modulated \halpha\ emission was consistent with an extended prominence system, characterized by an accumulation of hydrogen clouds near the corotation radius (in agreement with the previous work of \citealt{2001MNRAS.326.1057B}), and variations of the prominence system within a few days. Since this first  data set was not optimized to highlight fast changes, the purpose of the present study is to investigate in greater details the short-term variability of the prominence pattern, as well as the possible links of the coronal variability with the evolution of surface magnetic features, by investigating a denser time-series of observations.

We first present the new time-series of spectropolarimetric observations of \ap\ (Sect. \ref{sec:obs}), the reconstruction of brightness and magnetic field maps (Sect. \ref{sec:mapping}), our estimates of surface differential rotation from both \StI~and \StV~data (Sect. \ref{sec:DR}), and the tomography of \halpha~line profiles (Sect. \ref{sec:prom}). We then discuss our results with respect to previous works (Sect. \ref{sec:discuss}). In the final section, the main conclusions of this work are summarized (Sect. \ref{sec:conclusions}).

\section{Observations}
\label{sec:obs}


We carried out spectropolarimetric observations of V530 Per during four close nights (17, 18, 22, and 23 Oct 2018), using the 3.6-m Canada-France-Hawaii Telescope (Mauna Kea Observatory, Hawaii) equipped with the ESPaDOnS spectropolarimeter \citep{2006ASPC..358..362D}. The observational set up is identical to the one adopted by C20. ESPaDOnS is used in polarimetric mode, in which a spectral resolution of $\sim 65,000$ is achieved over a spectral range covering the whole optical domain (370 - 1050 nm). Under this mode, each polarimetric sequence is obtained from four independent subexposures with fixed 600~s exposure time, using different angles of two half-wave, rotatable Fresnel rhombs \citep{1993A&A...278..231S,1997MNRAS.291....1D}. Each sequence provides us with one circularly polarized (\StV) spectrum. We also use intensity (\StI) spectra reduced from the \StV\ subexposures to improve the temporal sampling in classical spectroscopy. We obtained in total 34 \StV~spectra, and 136 \StI~spectra  (see Tab. \ref{tab:obslog}). The raw images were automatically reduced, and normalized 1D spectra were extracted by the \texttt{Libre-ESpRIT} pipeline \citep{1997MNRAS.291..658D}. The number of exposures was not identical for every night, because of variable weather conditions. For most available observations, the peak signal-to-noise ration (S/N) of \StV~spectra reaches around 100, which is close to the level obtained with previous observations in late 2006. The weather on 22 Oct was slightly worse than the other nights, with a lower S/N (mostly below 100), while the following night was the best, with S/N levels larger than 106. All reduced spectra can be accessed through the PolarBase archive \citep{2014PASP..126..469P}.

The determination of the rotational cycle ($E$) for every observation is done according to the ephemeris of C20: 

\begin{equation}
\label{equ:emp}
\mathit{HJD}_{\mathrm{obs}} = 2454072.0 + 0.3205 \times E.
\end{equation}

\noindent Following the ephemeris, we can order the spectra according to their rotational phase and generate the dynamic spectra of Fig. \ref{fig:DS_stI_4night} and \ref{fig:DS_stV}, showing that the rotation coverage in each individual night was always over 50\%, and up to ~85\% during the last night. 
Given the adopted integration time, the phase smearing during the acquisition of a \StV~sequence is around $\sim10\%$ ($\sim2\%$ for \StI), which may reduce our sensitivity to the polarized signal generated by low latitude features, which have faster variations of their Doppler shift.

\section{Brightness and magnetic field mapping}
\label{sec:mapping}

Because of the similarity of the observational material used here and by C20, and to allow for a better comparison between the two studies, all modeling tools used for this work were strictly identical to C20, unless specifically stated. A summary of the various steps involved in the tomographic analysis is provided below.

\subsection{Least squares deconvolution}
\label{sec:lsd}

The spectral line profiles of V530 Per are heavily distorted, because of the very inhomogeneous surface brightness (recorded in \StI) and complex magnetic field geometry (in \StV). It is, however, especially difficult to study the profile shape in single lines of \ap, first because of the insufficient S/N, and also because of heavy blending generated by the rotational broadening. We take advantage, however, of the fact that all photospheric lines share a similar profile shape, with line-to-line differences mostly related to the line depth and wavelength (in \StI) or a combination of line depth, Land\'e factor, and wavelength (in \StV). To exploit this property and turn it into a statistical asset, we applied the so-called Least squares deconvolution (LSD) method to a list of photospheric lines \citep{1997MNRAS.291..658D}. Using this procedure, we obtain an average line profile, with a significantly increased S/N. 

The selected atomic line list was the nearest line list in the grid computed by \citet{2014MNRAS.444.3517M}, using the effective temperature \teff~$=5250K$, and an logarithmic gravity \logg~$=4.5$, making use of the same set of atmospheric parameters as C20. We ignored wavelength windows polluted by telluric or chromospheric lines, and finally picked a total of 5,726 lines. The LSD pseudo-profiles were obtained with velocity steps of 1.8 \kms\ and are produced with an equivalent Landé factor of 1.19 and an equivalent wavelength of 650~nm. An example of \StI~profile can be seen in Fig. \ref{fig:lsd_stI}, and all \StV~profiles are shown in Fig. \ref{fig:DS_stV}.

The dynamic spectrum (DS hereafter) of \StI~profiles illustrated in Fig. \ref{fig:DS_stI_4night} highlights a complex pattern of bumps and dips consistently repeated at different stellar rotation cycles.
These signatures, interpreted as the spectral imprint of dark and bright spots, generate trails progressively drifting in RV according to their rotational Doppler shifts. The thickest, positive trail close to the line center suggests the presence of a large, dark spot close to the visible rotational pole, but not exactly centered on the pole (which would not produce any variable Doppler shifts at all). A few smaller trails are visible superimposed on the main trail and several of them are also observed at higher Doppler velocities, which suggests the presence of smaller spots, located at lower latitudes. 

Although the patterns look almost the same for every night, a closer look reveals subtle differences between LSD profiles obtained during different nights, but at close-by phases (e.g., as shown in Fig.\ref{fig:lsd_stI}). These differences are consistently observed at other phases and obey to progressive radial velocity drifts of spectral signatures. They are likely the combined product of the surface differential rotation and any other type of intrinsic variability. As for \StV~profiles in Fig.\ref{fig:DS_stV}, we can also see structures repeatedly observed at close rotational phases. These features can be attributed to rotationally modulated Zeeman signatures produced by a complex magnetic field geometry.

\subsection{Zeeman-Doppler imaging}

From the time-series of LSD profiles, we reconstructed the surface brightness and large-scale magnetic field geometries of \ap\ with a ZDI code developed in Python and described in \citet{2018MNRAS.474.4956F}. The algorithm underlying this code is the one of \citet{2006MNRAS.370..629D}, which used the maximum entropy fitting routine of \citet{1984MNRAS.211..111S} for this ill-posed inverse problem, and a spherical harmonics decomposition of the magnetic field distribution. Assuming that the variations in the \StI~and \StV~spectra are primarily caused by the stellar rotation, we can reconstruct a Doppler imaging (DI) map of the stellar photosphere by using \StI~data, or a magnetic map with both \StI\ and \StV~data (the surface brightness distribution being taken as a prior input of the \StV~inversion). A simplified model of surface differential rotation, described in Sect. \ref{sec:DR}, is included in the inversion procedure. Owing to the very fast rotation of \ap\ that leads to a $\sim10\%$ difference in radius between the pole and the equator, the code also includes a Roche model to take into account the oblate shape of the star, as initially implemented by C20. 


The synthetic line profile produced by each surface element is modeled by a Gaussian function and is shifted according to its projected rotational velocity and scaled by the projection angle, a linear limb darkening function (e.g., \citealt{2005oasp.book.....G}) and a gravity darkening model (e.g., \citealt{1967ZA.....65...89L}). The limb darkening coefficient is interpolated from the table of \cite{2015A&A...573A..90M} (using the Kepler filter) and taken to be equal to $\eta = 0.73$. The gravity darkening coefficient $\beta =4b= 0.32$ adopted by C20 was an average value obtained for cool stars \citep{1967ZA.....65...89L}. We applied here a different value $\beta=0.46$, which is interpolated from the table of \citet{2011A&A...529A..75C} according to the fundamental parameters of \ap, and results in an equatorial brightness equal to $\sim 83\%$ of the polar one. As already stressed in C20, the $\beta$ coefficient has little impact on the resulting map.


All other input parameters of the tomographic inversion are equal to those discussed and adopted in C20. They include a projected rotational velocity equal to 106 \kms, and a stellar inclination angle equal to 40\degr. 

\subsubsection{Brightness map}



\begin{figure*}
    \centering
    \includegraphics[width=\textwidth,trim={4cm 7cm 3cm 4cm},clip]{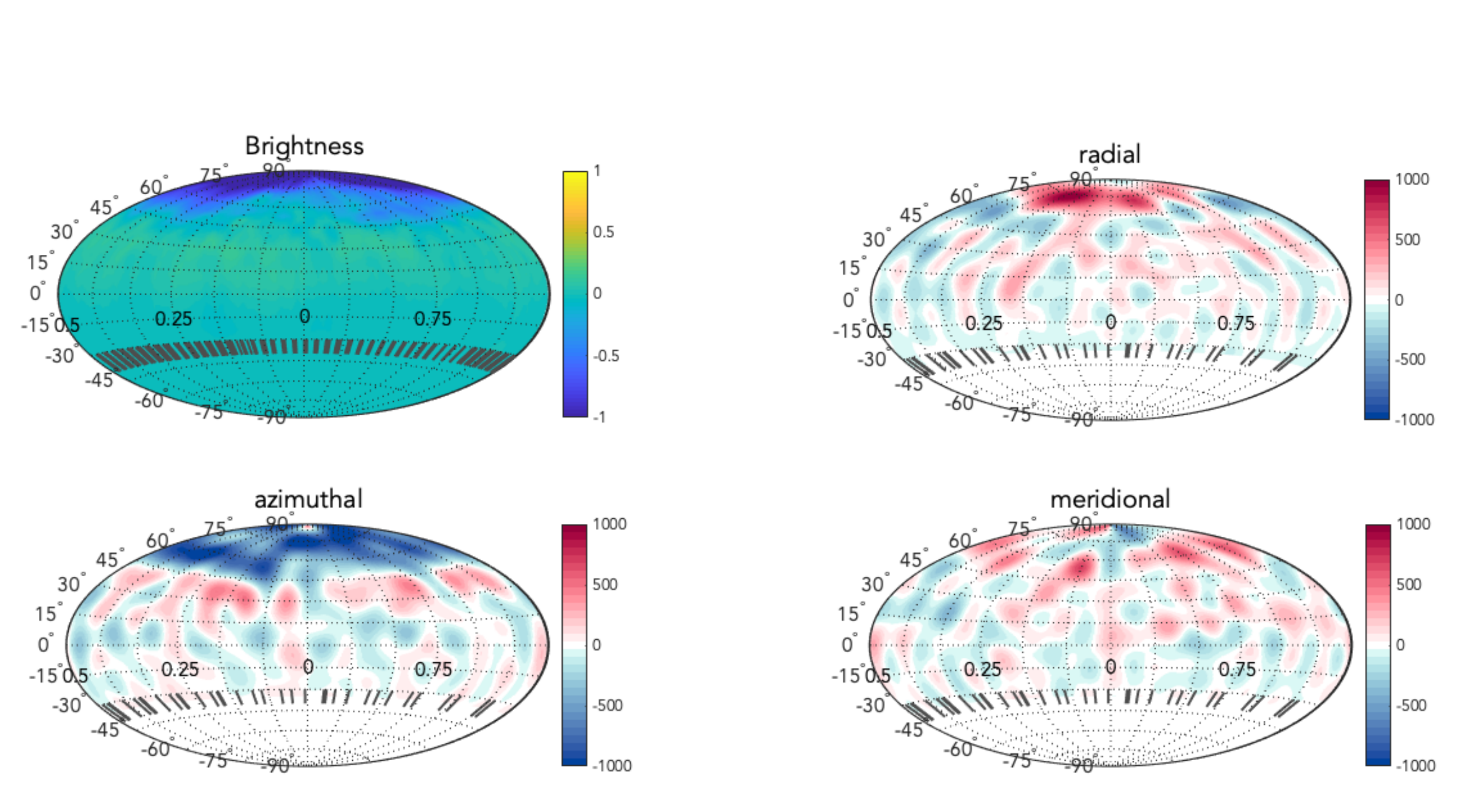}
    \caption{Reconstructed brightness and magnetic field maps of V530 Per in 2018. Top-left: Logarithmic brightness (normalized to the nonspotted brightness). For the sake of clarity, the Hammer projection was adopted and the gravity darkening was subtracted. Top-right, bottom-left, and bottom-right: radial, azimuthal, and meridional components of the magnetic field. The color scale illustrates the field strength, in Gauss. Meridional ticks at the bottom of the maps mark the rotational phases of our observations. The portion of the map below -40\degrr~of latitude is set to 0, as it is invisible to the observer. A polar view of the same maps can be found in Fig. \ref{fig:all_map_polar}.}
    \label{fig:all_map}
\end{figure*}{}

\begin{figure*}
    \centering
    \includegraphics[width=12cm,trim={0cm 6cm 0cm 6cm},clip]{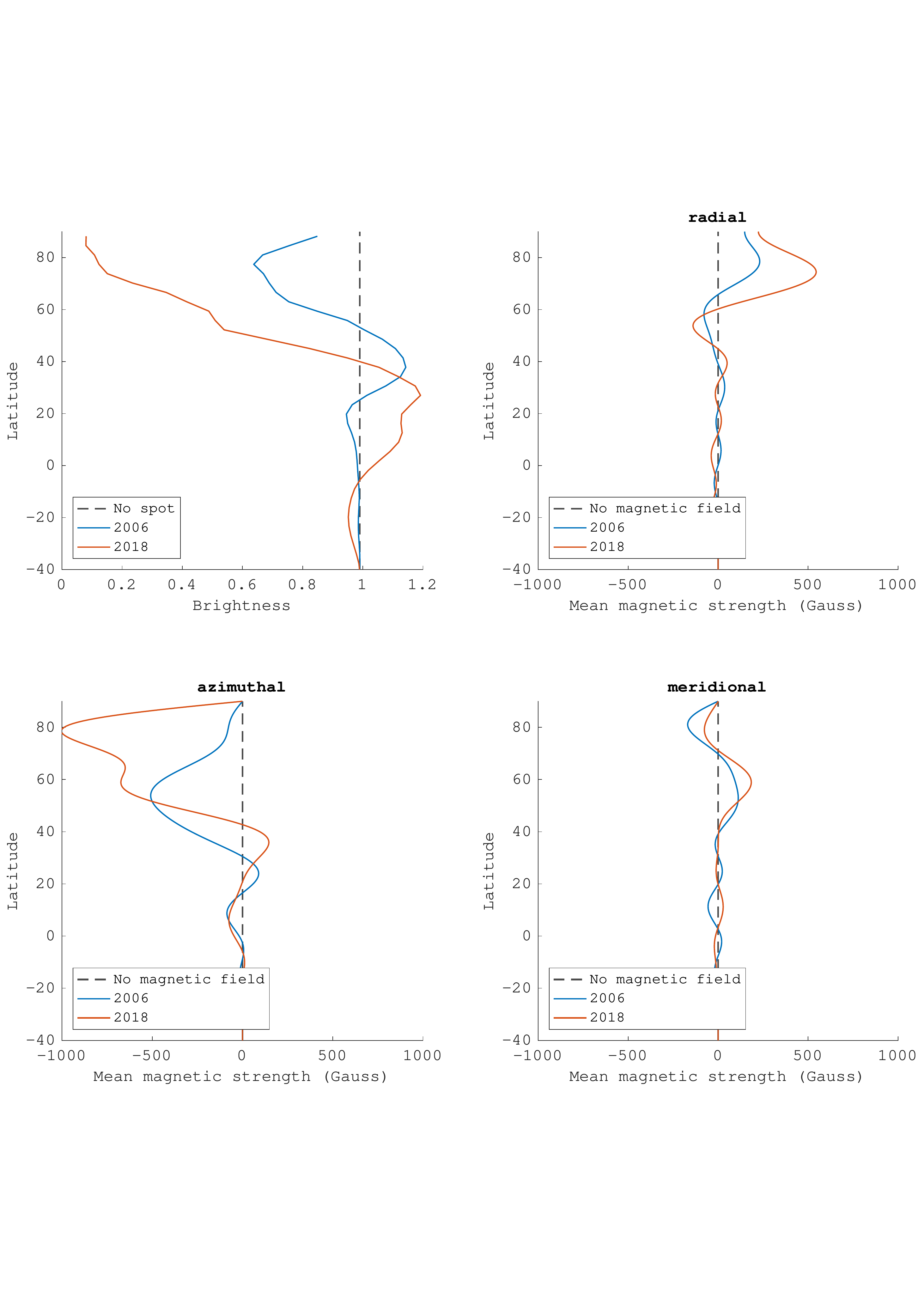}
    \caption{Normalized brightness (top left panel) and magnetic field components in spherical coordinates, as a function of stellar latitude. The blue and red lines show the data from 2006 and 2018, respectively.}
    \label{fig:bri_lat}
\end{figure*}

The overall stability, over several rotation periods, of activity signatures showing up in the \StI~DS in individual nights, led us opt to group all \StI~data together in the surface brightness reconstruction for a denser phase sampling. The inversion process also included the differential rotation parameters obtained in Sect. \ref{sec:DR}. The series of synthetic LSD profiles produced by the ZDI code is illustrated in Fig.\ref{fig:DS_stI_4night}, showing that the DI model is able to fit the majority of activity features, resulting in a reduced \chit\ (\chir\ hereafter) equal to 0.76. Although the level of residuals is negligible compared to the observed spectral signatures, there are some small features that the model cannot fully reproduce (e.g., the blue-shifted trail remaining at $\phi \sim 0.3 - 0.4$ on 23 Oct).
Since the DI algorithm tries to reproduce corotating brightness structures that do not evolve with time, except under the predictable shifts owing to differential rotation, the model residuals may be linked to the intrinsic evolution of the brightness tracers (changes in area, shape or intensity of the active regions). Given that the residuals are not consistent from one night to the next, the short lifetime of some surface structures may be responsible for this modest mismatch.

We note that the \chir\ we reached here is slightly larger than the one obtained with the 2006 data (which was equal to 0.55). This slightly degraded fit may be caused by a greater intrinsic variability of the spot distribution (emergence or decay of surface spots occurring faster and/or over larger areas) in the surface structures observed in 2018, bearing in mind that the \sn\ of both data sets is mostly the same, and that the time span of the new data is slightly shorter than in 2006. 

The most striking structure in the brightness map of Fig. \ref{fig:all_map} is a large, dark spot occupying the polar region. This prominent spot contrasts with the scarcity of smaller spots reconstructed at lower latitudes. The polar spot is centered close to a latitude of $\sim80$\degrr, and is slightly off-centered toward a phase between 0.3 and 0.5. Smaller and low contrast dark spots, separated from the polar spot, also show up at high to intermediate latitudes, down to $\sim45$\degrr. A predominance of bright features is observed from the equator to a latitude of about 40\degrr, as illustrated in Fig. \ref{fig:bri_lat} (see also the polar projection of the brightness map in Fig. \ref{fig:all_map_polar}, where the color scale was modified to highlight better the bright features). The same accumulation was reported by C20, although the bright features were structured in a series of more distinct individual spots in 2006, while they here take the shape of a nearly continuous belt, possibly due to a denser distribution of spatially unresolved spots. The total fraction of the surface covered by bright or dark spots $\mathcal{S}_\mathrm {tot}$ is equal to $\sim14\%$, as estimated by the following equation:

\begin{equation}
\centering
   \mathcal{S}_\mathrm {tot}=\frac{\sum_{i=0}^{n} |I_{i}-I_{0}|A_{i}}{\sum_{i=0}^{n} A_{i}} 
\end{equation}

\noindent where $I_{0} = 1$ is the unspotted brightness and $I_{i}$ is the brightness of the $i^\mathrm{th}$ pixel, of surface area $A_{i}$.

\subsubsection{Magnetic map}
\label{sec:map}

\begin{table}[ht]
    \centering
    \caption{Magnetic field characteristics of V530 Per in 2018 and 2006 (2006 values taken from C20).}
    \begin{tabular}{llll}
        \hline
       & Parameter & Value (2018) & Value (2006) \\
        \hline
a&        $\langle B \rangle$    & 222 G & 177 G         \\
b&        $|B_{peak}|$           & 1616 G & 1088 G      \\
c&        toroidal               & 68  \% (tot)  & 64 \%          (tot)\\
d&        axisymmetric           & 65 \%(tot) & 53 \%            (tot)\\
e&        pol axisymmetric  & 36 \%  (pol)& 16 \%            (pol)\\
f&        tor axisymmetric  & 79 \%(tor) & 74 \%            (tor)\\
g&        dipole                 & 6.3 \%  (pol)& 1.2 \%           (pol)\\
h&        quadrupole             & 6.6 \% (pol) & 3.3 \%          (pol)\\
i&        octopole               & 7.3 \% (pol) & 5.4 \%          (pol)\\
 j&  pol $\ell > 3$ & 79.3 \% (pol) & 90.1 \% (pol)\\  
k&        tor $\ell = 1$    & 2.6 \% (tor)& 8 \%  (tor)\\
l&        tor $\ell = 2$    & 8.6 \%(tor) & 21 \% (tor)\\
m&        tor $\ell = 3$    & 13.7 \% (tor)& 20 \% (tor)\\
n&        tor $\ell > 3$    & 75.1 \% (tor)& 51 \% (tor)\\
        \hline
    \end{tabular}
    \begin{tablenotes}
      \small
      \item \textbf{\tt{Abbreviations:}}~ tot=total, pol=poloidal, tor=toroidal
      \item \textbf{\tt{Note:}}~The quantities listed include (a) the average magnetic field strength $\langle B \rangle$, (b) the unsigned peak magnetic field strength $|B_{peak}|$, (c) the ratio of toroidal field energy with respect to the total magnetic energy, (d) the ratio of magnetic energy in axisymmetric modes ($m = 0$) over the total energy, the same quantity but limited to the poloidal (e) and toroidal (f) magnetic component, the ratio of the dipole, quadrupole, octopole, and $\ell > 3$ (g, h, i, j) as a fraction of the poloidal component, and ($\ell=1,2,3,>3$) subcomponents of the toroidal field energy, as a fraction of the toroidal field energy (k, l, m, and n).
    \end{tablenotes}

    \label{tab:magE}
\end{table}

Similarly to the brightness inversion, the magnetic field reconstruction benefited from the dense phase coverage of the set of \StV~LSD profiles (Fig. \ref{fig:DS_stV}), which is especially critical here since the Zeeman signatures are barely detected in individual \StV\ LSD profiles. Owing to the large relative noise, it is also impossible to readily notice, with the naked eye, any changes in the profiles that could be attributed to a surface shear. A search for signatures of differential rotation was however performed (Sect. \ref{sec:DR}), and the resulting parameters were used in the field reconstruction. The model fit the data with a reduced \chit~of 0.92, again slightly larger than the one achieved with the 2006 data (\chit~of 0.9). The magnetic field model includes spherical harmonics modes up to $\ell = 15$. The resulting map, shown in Fig. \ref{fig:all_map}, unveils a complex surface distribution of magnetic fields. Similarly to C20, the small structures are also reconstructed, but with a slightly reduced strength, if we adopt a slightly larger \chit~of 1. For the radial field component, the strongest and most visible structure is a positive magnetic spot close to the polar region. Its peak value is in excess of 1~kG, at a latitude of $\sim75$\degrr~and a rotation phase of $\phi\sim 0.2$. The main structures visible in the azimuthal field component is of negative polarity, and occupies the whole region at latitudes greater than $\sim45$\degrr. Apart from these two distinctive magnetic features, at lower latitudes a complex distribution of spots is reconstructed in the radial, azimuthal, and meridional field components, with no obvious counterpart in the brightness map.

We estimated a series of magnetic energy parameters from the modeled spherical harmonics coefficients, detailed in Tab. \ref{tab:magE}. There is a large difference between the unsigned average magnetic field strength $\langle B \rangle$, and the unsigned peak magnetic field strength $|B_{peak}|$, which is a consequence of the complex field structure, exhibiting strong magnetic fields in localized spots. Simply considering that the magnetic energy is proportional to $B^2$, the toroidal field component stores most of the energy ($\sim68\%$) and has a mostly axisymmetric structure ($\sim 79\%$ of its energy shows up in modes with $m = 0$). The poloidal field component displays a lower axisymmetry ($\sim~36\%$ of the energy with $m = 0$). For both the toroidal and poloidal field, most of the energy is stored in the high order spherical harmonics components ($\ell>3$), which is another way to illustrate the complexity of the large-scale magnetic field. As a rough estimate of the uncertainty on these values, we varied the input stellar parameters (\vsini, the inclination angle, the equatorial rotation period, and the surface shear) over their confidence interval, as well as the target \chit. We conclude that the stellar parameters have little effect on the estimated magnetic energy ($\sim2\%$), except \vsini\ that is able to modify the magnetic values by up to 10\%. Changing the \chit\ within a reasonable range can also modify the derived magnetic characteristics by $\sim10\%$.

\section{Differential rotation}
\label{sec:DR}

\begin{figure}
    \centering
    \includegraphics[width=8cm]{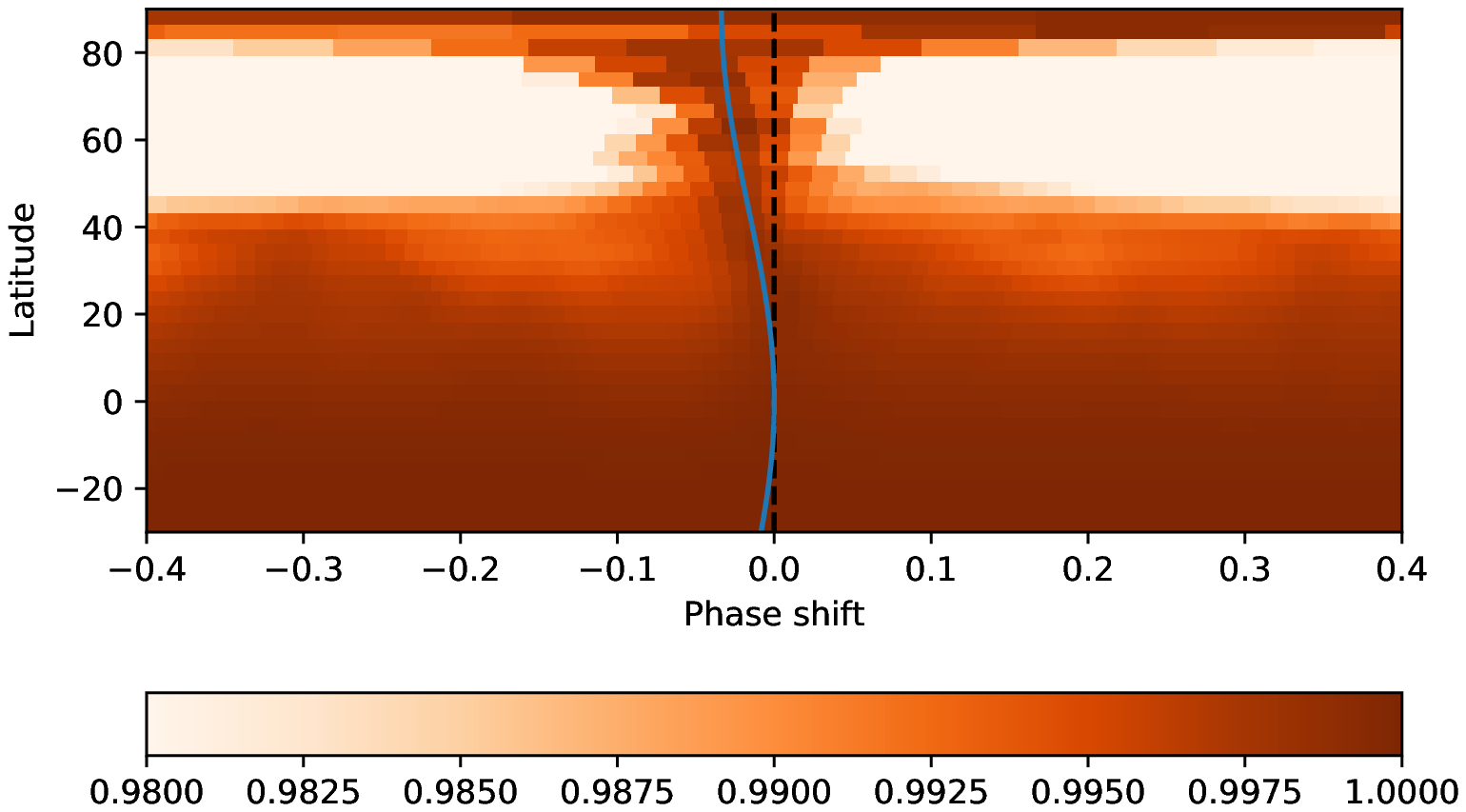}
    \includegraphics[width=8cm]{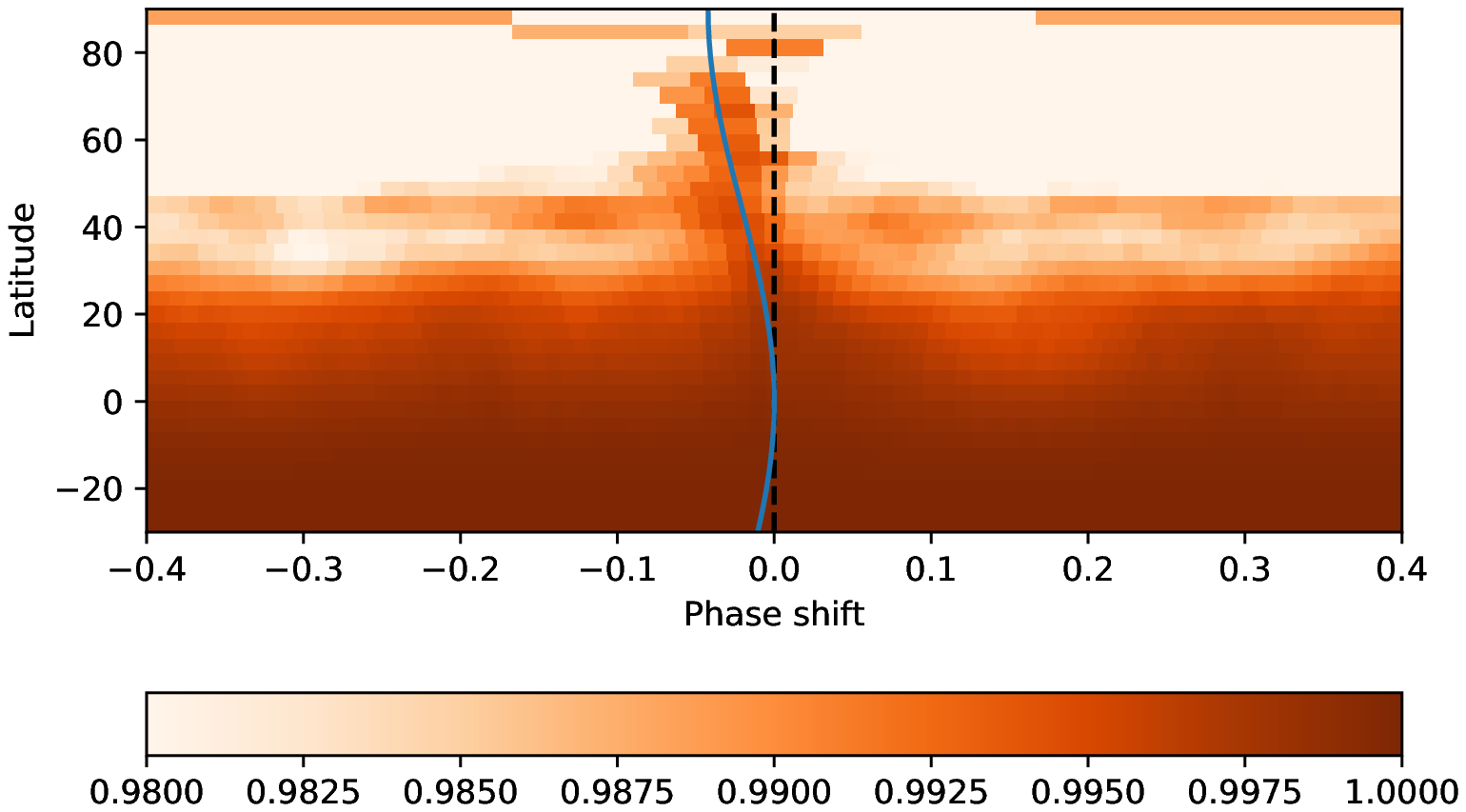}
    \caption{Top: cross-correlation map obtained by comparing two brightness maps obtained using 2018 data from 18 Oct, versus 23 Oct (with a gap of 15.6 rotation periods between the two maps). Bottom: same, but for the two observing nights of our 2006 data (18.7 rotation periods apart). The blue lines show the differential rotational law derived using the sheared imaged method in Sect. \ref{sec:DR} (top panel), and in C20 (bottom panel). The black, dashed line marks a null phase shift (which is equivalent to a solid-body rotation).}
    \label{fig:ccr}
\end{figure}

\begin{figure}
    \centering
    \includegraphics[width=8cm,trim={1.5cm 0cm 1.5cm 0cm},clip]{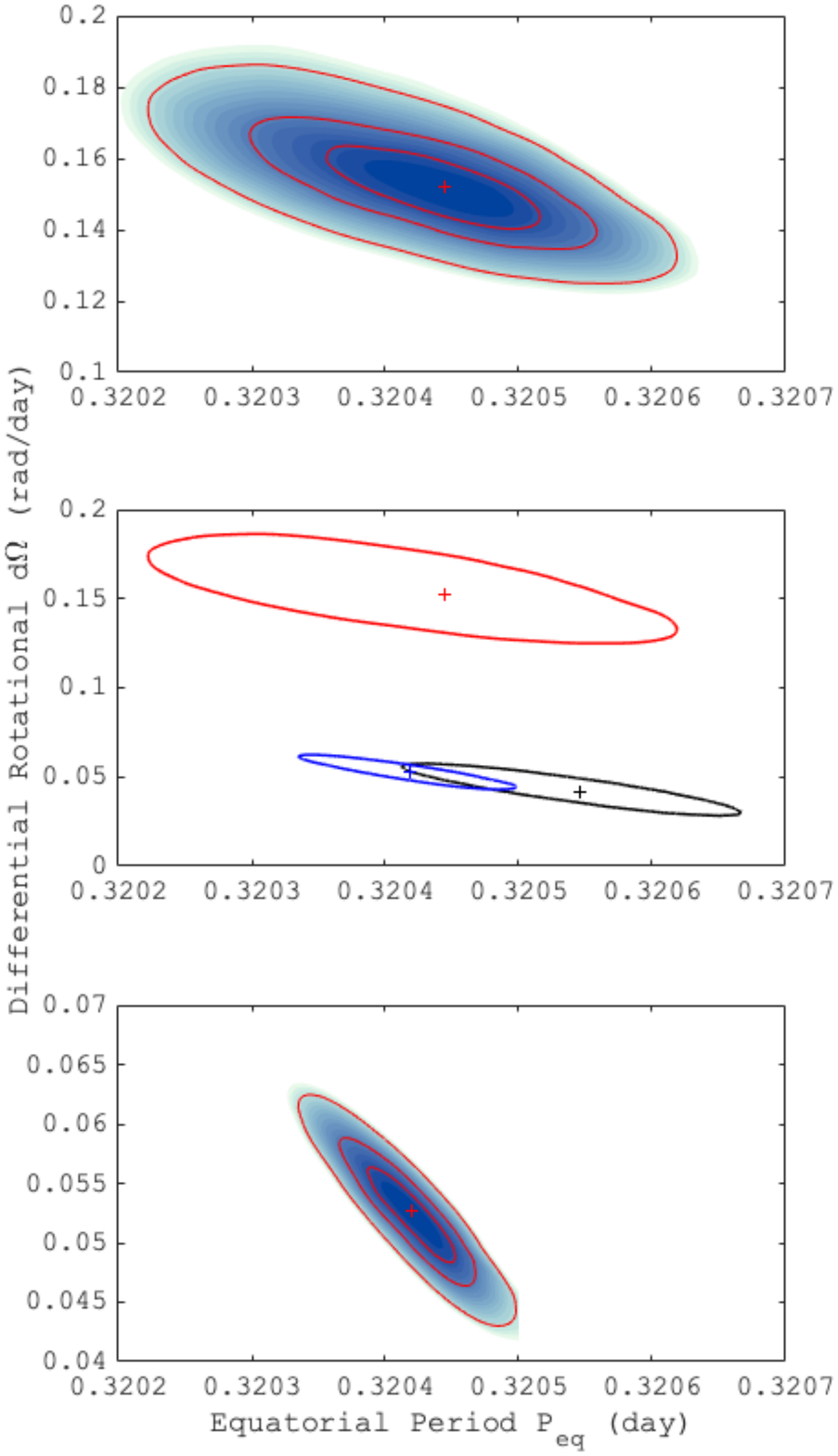}
    \caption{Reduced \chit\ maps for the equatorial rotational period $\Omega_\mathrm{eq}$ and differential rotation $d\Omega$ for \StI~(bottom panel)  \& \StV~data (top panel). The three, red solid lines illustrate the $1\sigma$, $2\sigma$, and $3\sigma$ confidence intervals. Middle panel: a comparison of 3$\sigma$ regions for \StI~in 2018 (blue), \StI~in 2006 (black), and \StV~(red). Crosses mark the location of peak values.
  }
    \label{fig:dr}
\end{figure}

The subtle changes seen in the intensity line profiles, at nearby phases repeatedly observed over the course of our observing run (Fig. \ref{fig:lsd_stI}), suggest that the brightness distribution is changing with time. In this section, we investigate whether a fraction of its variability can be modeled under the assumption of a differentially rotating surface. As an initial test of this idea, we compare two brightness maps (not shown here) reconstructed from data obtained on Oct 18 and 23, respectively. This specific choice of dates is a compromise between a sufficiently large temporal lever arm, and a good phase coverage. The comparison is performed as a cross-correlation of the two maps (Fig. \ref{fig:ccr}), following \cite{1997MNRAS.291....1D}. We observe that the surface structures are systematically shifted in phase between the two dates, and that the shift increases with the latitude. Most of the usable cross-correlation signal is seen at latitudes greater than about 40\degrr, because of a lack of surface brightness tracers closer to the equator. In the same figure, we show the same approach applied to the data of C20, leading to very similar conclusions. The blue lines display a simple solar-like surface shear law (see below), showing that this simple description of the shear is consistent with our observations. 

As a second step, we used the built-in sheared image ZDI method \citep{2000MNRAS.316..699D,2002MNRAS.334..374P}, where a solar-like differential rotation is implemented as part of the ZDI model, following a simple solar-like prescription:

\begin{equation}
\label{equ:DR}
\Omega(\theta) = \Omega_{\mathrm{eq}} - d\Omega \sin^{2}\theta
\end{equation}

\noindent where $\Omega_{\mathrm{eq}}$ is the rotation rate of the equator, $d \Omega$ the pole to equator gradient in rotation rate, and $\theta$ the latitude.

We used the same ZDI model parameters as those discussed in Sect. \ref{sec:mapping}, and carried out a grid search by varying $\Omega_{\mathrm{eq}}$ and $d\Omega$. We choose here to fix the entropy of the model, so that the output is a \chir~landscape, over a range of parameter values (Fig.\ref{fig:dr}). 

For \StI, the peak value is located at an equatorial period $P_\mathrm{I}= 0.32042 \pm 0.00005$ d, and a shear d$\Omega_\mathrm{I} = 0.053 \pm 0.004$ rad.d$^{-1}$, in overall agreement with C20. While a previous search for differential rotation using \StV\ was inconclusive with the 2006 observations, the much denser phase coverage in 2018 led to a detection, with the best parameters equal to $P_\mathrm{V}=0.32045\pm 0.0001$~d, and $d\Omega_\mathrm{V} =0.15 \pm 0.01$ rad.d$^{-1}$. The $d\Omega$ value derived from the large-scale magnetic field is, therefore, $\sim3$ times larger than the value estimated from surface brightness. The shear search in \StV\ was performed using the brightness map as a prior, implying that the brightness was sheared by the same differential rotation parameters as \StV. We repeated the differential rotation search for \StV\ but assumed a constant surface brightness (as done in most previous ZDI measurements), and found a shear value within error bars of our first \StV\ estimate.

\section{Prominences}
\label{sec:prom}

\begin{figure*}
    \centering
    \includegraphics[width=12cm]{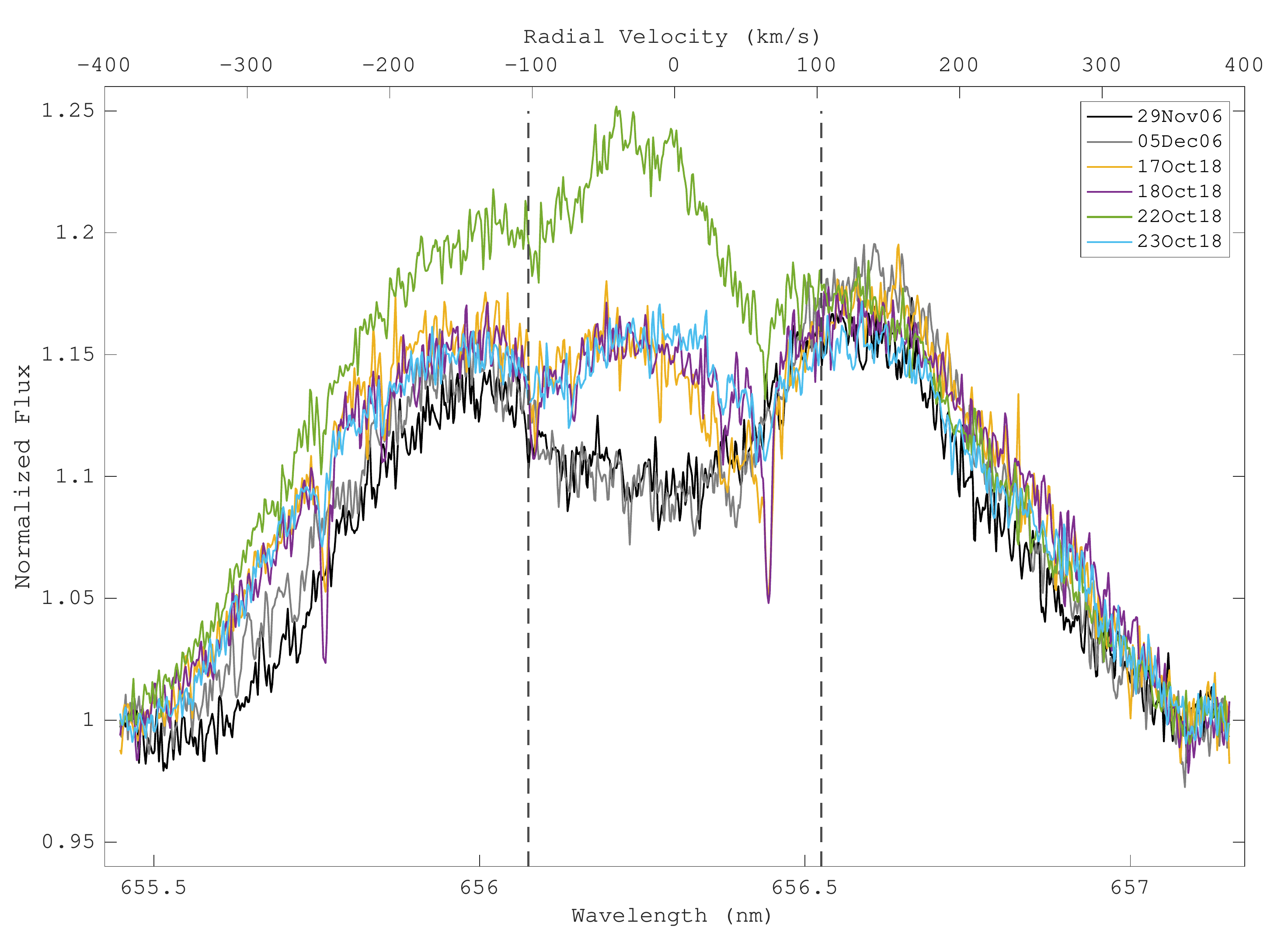}
    \caption{\halpha~profiles averaged over individual nights in 2006 and 2018.  Two vertical dashed lines show the $\pm$\vsini\ radial velocities.}
    \label{fig:mean_halpha}
\end{figure*}

\begin{figure*}
    \centering
    \includegraphics[width=18cm,trim={0cm 0cm 0cm 0cm},clip]{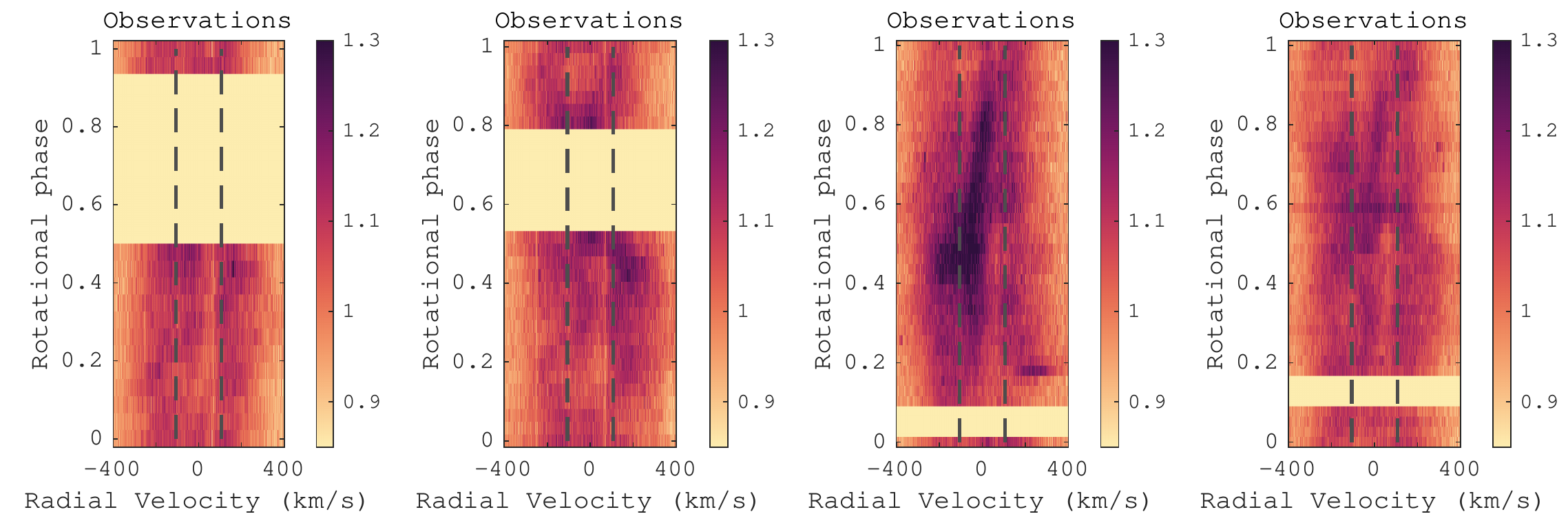}
    \caption{Dynamic spectra of \halpha\ line profiles, phased according to the stellar rotation period. From left to right, the panels show the data of 17, 18, 22, and 23 Oct 2018. Rotational phases are computed according to Eq. \ref{equ:emp}. Vertical dashed lines show $\pm$~\vsini.}
    \label{fig:DS_halpha_obs}
\end{figure*}

\begin{figure*}
    \centering
    \includegraphics[width=8cm,trim={3cm 8cm 3cm 8cm},clip]{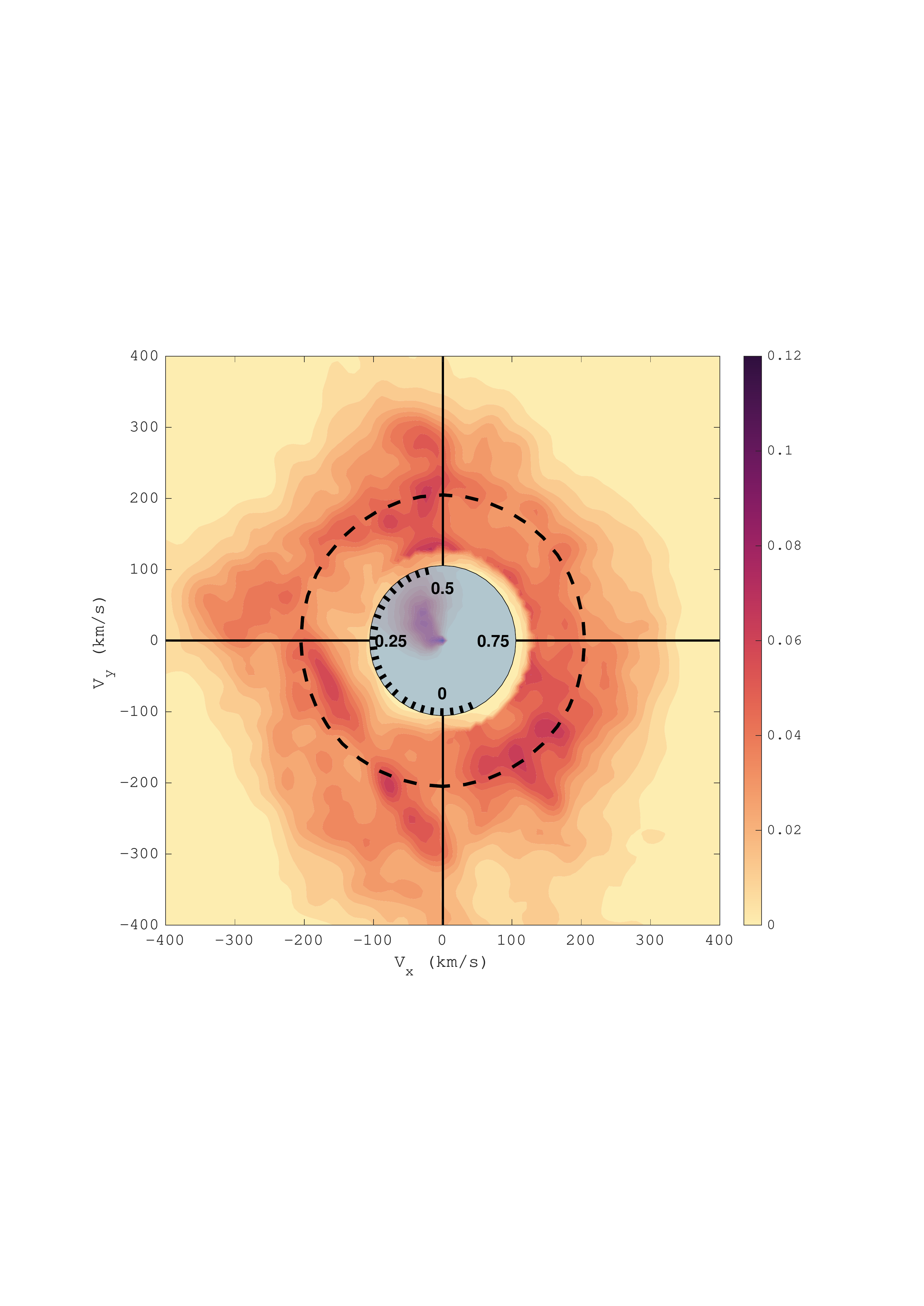}
    \includegraphics[width=8cm,trim={3cm 8cm 3cm 8cm},clip]{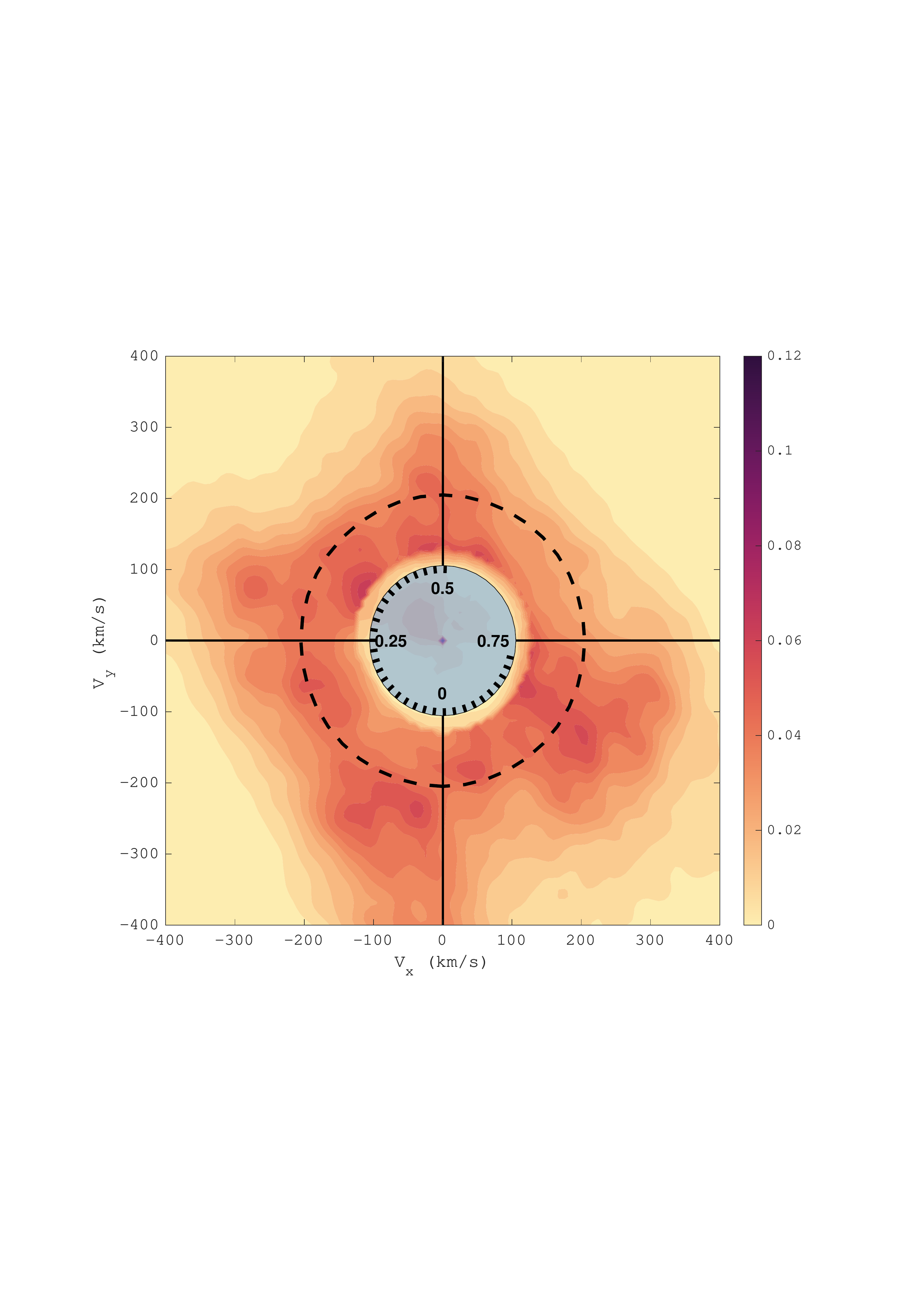}
    \includegraphics[width=8cm,trim={3cm 8cm 3cm 8cm},clip]{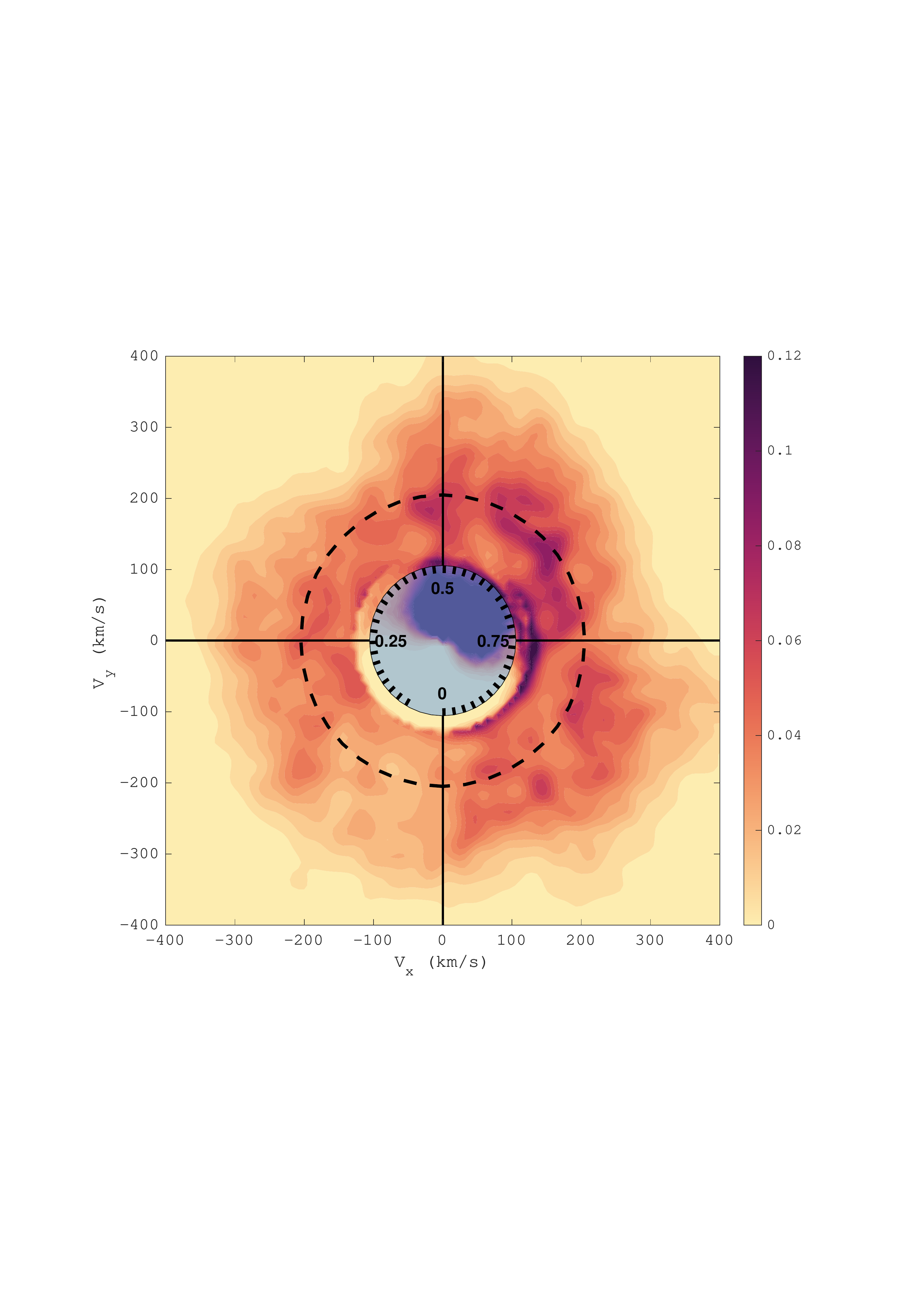}
    \includegraphics[width=8cm,trim={3cm 8cm 3cm 8cm},clip]{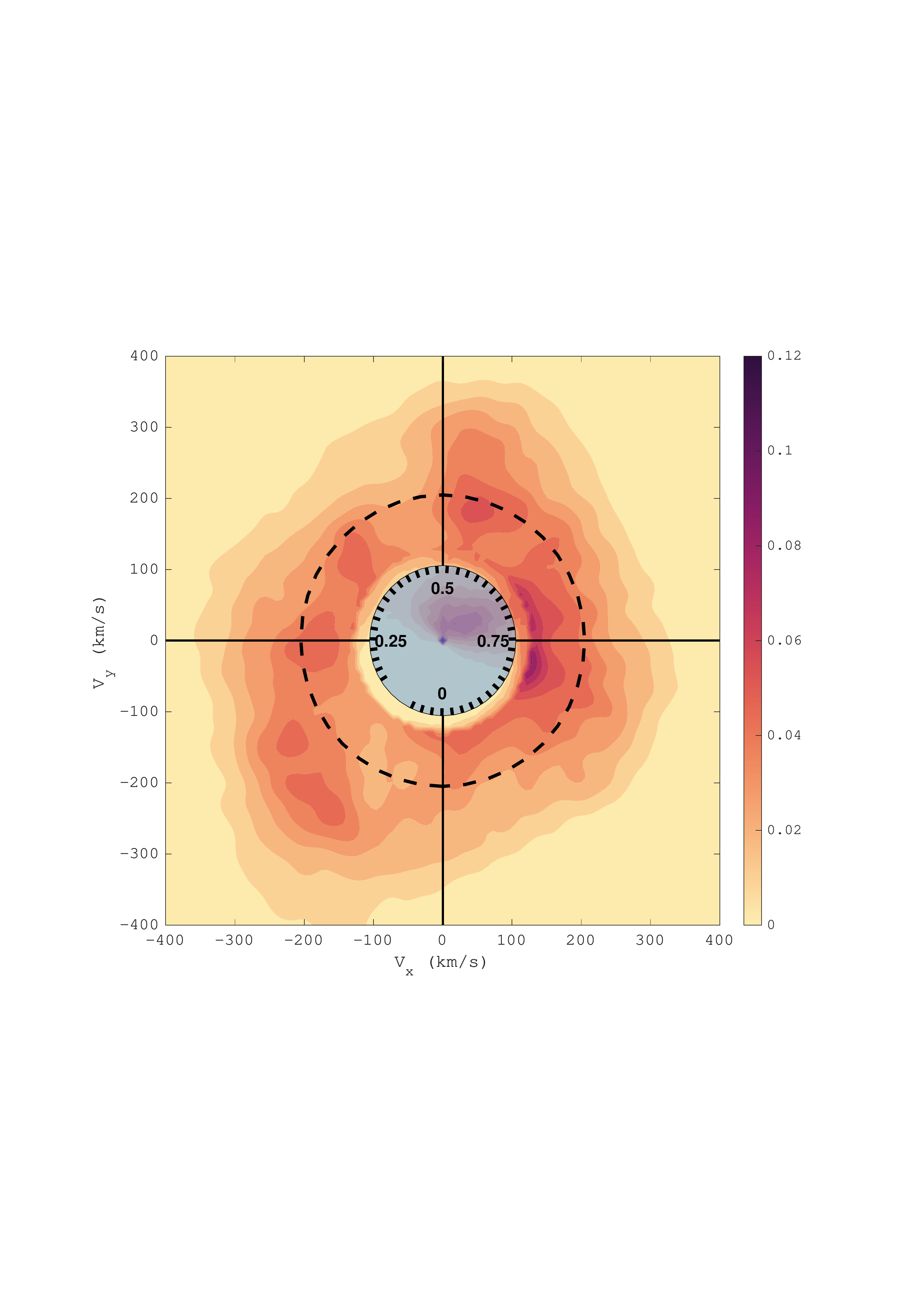}
    \caption{Prominences Map of V530 Per in 17 (upper-left),18 (upper-right), 22 (lower-left),23(lower-right) Oct. The inner, filled blue circle represents the stellar surface. Radial ticks inside this circle give the rotational phases of \halpha\ observations. The outer dashed circle is the corotation radius. The color scale depicts the local \halpha~equivalent width, in units of picometers per 8 km~$\rm{s}^{-1}$ square pixel.}
    \label{fig:prom_map}
\end{figure*}
The \halpha~line is observed in emission in all of our observations. In this section, we investigate the short-term evolution of this spectral feature within individual nights and over the full duration of our observing run.

When all observations of a single night are averaged together, a triple-peaked profile is observed (Fig. \ref{fig:mean_halpha}), as opposed to the 2006 data where a double peak was repeatedly reported. Both side peaks show up at around $\pm150$\kms\ from the line center (in 2006 and 2018), while the central peak is close to the line center. The \halpha~profile of Oct 22 stands out, with a distinctly larger amplitude of the central and blue-shifted peaks. The other nights display similar profiles, and the amplitude of their red and blue peaks tend also to agree with the 2006 observations.  

The evolution of \halpha\ within individual nights is illustrated as a set of dynamic spectra in Fig. \ref{fig:DS_halpha_obs}. A fraction of the observed variability is consistent with the rotational modulation of hydrogen clouds trapped in corotation with the star, since part of the observed emission is stable over more than one day, when phased according to stellar rotation cycles given by Eq. \ref{equ:emp}. This is shown by, for example, similarities in the emission patterns observed on 17 and 18 
Oct (three rotation cycles away from each other), especially at rotational phases smaller than 0.4. According to the estimate of C20, the coronal large-scale magnetic field of V530 Per is able to trap and support large prominences, which we assumed to be responsible for this rotationally modulated \halpha~emission. 

The relatively stable \halpha\ configuration observed during the first two nights is, however, considerably different from the one observed on 22 Oct. After three nights without observations, several additional emission components are observed, resulting in an overall increase of the emission for this specific night. We first note a broad emission peak around phase 0.5 and at negative velocities (between roughly -200 \kms\ and 0 \kms). This line bump was not so prominent during the two previous nights (if it was present at all), and it becomes much weaker again during the last observing night. Another emission component takes the shape of a trail close to line center and extending from phase $\phi\sim0.4$ to phase $\phi\sim0.9$. It is not easy to decide whether this trail was already there in the previous observing nights, owing to the incomplete phase sampling. It is, however, present in the fourth night, although with a much reduced brightness. This trail (which is responsible for the central peak seen in Fig. \ref{fig:mean_halpha}) is confined within a range of velocities going from $\sim-50$~\kms\ to $\sim+30$~\kms. The repeated observation over consecutive nights, at similar phases, definitely shows that this spectral feature is rotationally modulated. It is visible during only a fraction of the rotation period, suggesting that it is eclipsed behind the star during part of the rotation cycle. This is consistent with a hot chromospheric spot located at intermediate latitudes (since a prominence would more likely be seen as an absorption feature when transiting in front of the stellar disk). Finally, a burst-like event takes place at phase $\phi\sim0.18$ on 22 Oct and with a blue shift of $\sim +230$~\kms, with a lifetime shorter that our temporal resolution (i.e. $< 600$ sec).

Using this material, we followed the tomographic procedure described by C20 to reconstruct the velocity distribution of the prominences, using individual nights data. The Doppler tomography inversion uses the code of \citet{2000MNRAS.316..699D}, inspired from the tomographic approach developed by \citet{1988MNRAS.235..269M} or \cite{1996MNRAS.281..626S}. In this simplified model, we assume that the \halpha~emitting material is optically thin, and corotates with the star. A description of this simple model can be found in \citet{2000MNRAS.316..699D} and C20.

The modeled DS are shown in the middle panels of Fig. \ref{fig:DS_halpha}, illustrating that the tomographic inversion is able to fit a majority of the observed features. We reached, however, a larger \chit\ ($\sim9.5$, with values ranging from 9 on 17 Oct to 10 on 22 Oct)~than the one obtained from the 2006 data (\chit$\sim8$). This slightly degraded fit suggests that our simple model is challenged by these observations, either because some level of non-rotational variability affects our data over individual nights, or because some basic assumptions of the model are not consistent with the data (e.g., the fact that the emitting clouds are not supposed to be eclipsed by the star). The model residuals shown in the lower panels of Fig. \ref{fig:DS_halpha} are generally an order of magnitude lower than the observed emission, which suggests that most of the periodic patterns were successfully reconstructed by the tomographic code. Some of the phases and velocities displaying a significant mismatch with the model correspond to burst-like events described above (structures that are too short-lived to follow a rotationally-modulated pattern). For instance, at $\phi\sim0.18$ on 22 Oct, the peak on the right of the line profile cannot be reproduced due to its very brief appearance in our data. Another example of mismatch is the trail seen close to the line core on 22 and 23 Oct. In this second case, where rotational modulation seems at play, the poor fit is likely owing to the intermittent visibility of the cloud, that spends part of the rotation cycle hidden behind the stellar disk.

Apart from the fast changing component of \halpha\ emission that escapes our modeling attempts, most of the prominence pattern is correctly fit for all four nights. In the resulting maps, the emitting material accumulates around the corotation radius ($\sim1.9~R_*$), and forms an extended ring-like structure. The reconstructed pattern inside the \vsini~limit is confined within a relatively small phase interval, centered around phase $\phi\sim0.4$ during the first two nights, then showing up around phase $\phi\sim0.7$ during the last two nights, with a prominent emission peak on Oct 22 in relation to the emission trail observed close to the line center. 

We varied the rotation period in our tomographic model, and identified a preferred period of 0.37~d on 17 Oct. This estimate is consistent with previous findings of C20, who reported that their inversion was optimized with a 0.36-0.39~d period (depending on the night), possibly owing to a less effective corotational locking of prominences at large distances from the star. The period search was inconclusive for all other nights in 2018, possibly because of the fast variability hiding the rotational modulation. 

\section{Discussion}
\label{sec:discuss}

\subsection{Surface brightness and magnetic field}

Large-scale brightness and magnetic field geometries derived in this work show some clear similarities with the maps presented by C20. At both epochs, the brightness distribution was dominated by a dark spot anchored at high latitude. The second feature recognizable in both maps is an accumulation of bright spots at intermediate latitudes. Both maps display a spot coverage slightly larger than 10\%. This consistent latitudinal dependence of the brightness is illustrated in the upper-left panel of Fig. \ref{fig:bri_lat}. Beside the global consistency of the two maps, a clear evolution is seen regarding the high latitude spot, which is much darker at the new epoch, with a minimum normalized brightness decreasing from $\sim0.7$ in 2006 to $\sim0.25$ in 2018. The shape and location of this giant spot varied as well, from a location that did not cover the pole in the Doppler map of \cite{2001MNRAS.326.1057B} and in our 2006 data, to a nearly centered spot in 2018. Such a clear evolution in the axisymmetry of the main polar spot was not reported in other young solar-type stars with long-term monitoring (e.g., AB Dor, \citealt{1997MNRAS.291....1D,2003MNRAS.345.1187D}). We also note that the latitude of maximum brightness was shifted by approximately 10\degr\ towards lower latitudes in 2018.

Similarly to the larger spot coverage observed close to the pole, we reconstructed stronger magnetic fields at high latitudes, for both the radial and azimuthal field components (Fig. \ref{fig:bri_lat}). The latitude where the radial field strength is maximal is roughly the same in 2006 and 2018, while the latitude of maximal (unsigned) azimuthal field has been shifted by about 20\degrr towards the pole. While the unsigned azimuthal field was maximal close to the limit of the polar spot in 2006, in 2018 the azimuthal component was strong well inside the dark polar structure (Fig. \ref{fig:all_map_polar}). The field strengths of the latitudinal maxima of the radial and azimuthal field components roughly doubled from 2006 to 2018, with no observed polarity reversal. 

The 2006 and 2018 magnetic geometries translate into relatively similar distributions of the magnetic energy (Table \ref{tab:magE}). More than half of the energy was contained in the high order components ($\ell$\gt3) for both poloidal and toroidal field. We note that the reconstructed magnetic field is more axisymmetric in 2018 ($\sim$65\% of the energy in modes with $m = 0$) than in 2006 ($\sim$53\%), and the majority of this variation can be attributed to the poloidal axisymmetric component (with an increase from $\sim$16\% to $\sim$36\%). The average magnetic field strength obtained in 2018 ($\langle B \rangle \sim 222$~G) and the unsigned peak magnetic field strength ($|B_{peak}|\sim 1616$~G) are both larger than in 2006 (where 177~G and 1088~G were measured, respectively). The overall increased magnetism observed in 2018 suggests a variable activity level, but the very scarce available monitoring makes it impossible to conclude about the long-term nature of this variability, which may be mostly chaotic as reported for other young, rapidly-rotating dwarfs like AB Dor or LQ Hya \citep{2003MNRAS.345.1145D}.

\subsection{Differential rotation}

The progressive, latitude-dependent phase shift of brightness spots can be convincingly approximated by a solar-like differential rotation law, as illustrated in Fig. \ref{fig:ccr} where a simple cross-correlation of latitudinal strips is calculated, without any prior on the shear law. Using the same methodology, a solar-like shear pattern was also found consistent with successive Doppler maps of other young, rapidly rotating stars, e.g., the K dwarfs AB Dor \citep{1997MNRAS.291....1D} and LO Peg \citep{2005MNRAS.356.1501B}, the young G dwarf HD 141943 \citep{2011MNRAS.413.1939M}, and the post T Tauri star LQ Lup \citep{2000MNRAS.316..699D}. Recent reports on GJ791.2A and GJ 65 AB \citep{2017MNRAS.471..811B}, again based on the cross-correlation approach, suggest that very active M dwarfs can also follow a solar-like differential rotation law. A more complex latitudinal rotational dependence, similar to a Jupiter-like pattern, was suggested by numerical simulations at very fast rotation rates \citep{2017ApJ...836..192B}. The smoother dependence observed for \ap\ may suggest that zonal flows would require even shorter rotation periods (which, in practice, would make this phenomenon fairly marginal among young Sun-like stars), or that any departures from a simple solar law are sufficiently subtle to remain hidden in the noise. 

We identified a higher shear level using \StV~profiles ($\sim0.15$~rad/d, $\sim3$ times larger than the \StI\ estimate). Similar results were repeatedly obtained for the young dwarfs AB Dor and LQ Hya \citep{2003MNRAS.345.1145D}, and the T-Tauri stars Par 2244 \citep{2017MNRAS.472.1716H} and V410 Tau \citep{2019MNRAS.489.5556Y}. Assuming that this observation is not an artifact of the inversion procedure, it has been suggested by \cite{2003MNRAS.345.1145D} that such differences may be linked to Stokes I and V tracing different depths within the star (depending on how deep the surface brightness and magnetic regions are actually generated), which is an interpretation also proposed for the Sun \citep{2000SoPh..191...47B}. We also note that the \StI\ shear measurement obtained in 2018 is consistent within 3$\sigma$ with the one obtained in 2006, while temporal changes in this value were previously reported for AB Dor \citep{2002MNRAS.329L..23C}.

\subsection{Short-term variability of the prominence distribution}

Similarly to the previous observations of C20, the \halpha~emission of V530 Per in 2018 shows signatures of rotational modulation (Fig. \ref{fig:DS_halpha_obs}), suggesting that most of the prominence system is forced to corotate with the star. Two emission peaks show up at a similar velocity of about $\pm150$\kms\ at both epochs (Fig. \ref{fig:mean_halpha}), with similar flux levels. We attributed them to the large hydrogen clouds accumulated around the corotational radius and trapped by coronal magnetic loops, as already proposed in
other fast-rotating stars \citep{1989MNRAS.236...57C,1989MNRAS.238..657C, 1996MNRAS.281..626S, 2000MNRAS.316..699D}.

The central emission peak was not observed in 2006. In 2018, it is linked to a trail in the DS that shows evidence of rotational modulation, and remains confined at relative radial velocities smaller than about 50 \kms. This peak is responsible for the features inside the \vsini\ limit in Fig. \ref{fig:prom_map}. The small Doppler shift, combined with the eclipse of this signal during about 30\% of the rotation, suggests that it is produced by a hot chromospheric point anchored at intermediate latitudes. 

In the DS of 22 and 23 Oct, this central trail in \halpha\ evolves in phase with the main, positive trail in the \StI~DS. This phase correlation suggests that this bright, short lived chromospheric feature is lined up with the phase of the off-centered polar spot. A similar observation was  reported for the K dwarf RE 1816+541 \citep{1998A&A...337..757E}, and the weak line T Tauri star TWA 6 \citep{2008MNRAS.385..708S}. 

The observed differences in the shape and intensity of the \halpha\ line between successive nights suggest a day-to-day variation of the prominence arrangement. The emission recorded at the corotating radial velocity is less dramatically affected by these fast changes than the central emission peak. Surface activity tracers do not obviously reflect this rapid evolution, as illustrated by the \StI\ data of Fig. \ref{fig:DS_stI_4night}, or by the \StV\ profiles of Fig. \ref{fig:DS_stV}. The reconstruction of a series of brightness maps using data of individual nights (not shown here) does not unveil any noticeable changes in the spot pattern, or at least not at a level that can be safely trusted as a genuine evolution (versus spurious differences owing to the different phase coverage, for instance). The reconstruction of magnetic maps for individual nights increases the noise contribution, which has the effect of hiding even more efficiently any possible variability. The only measured source of surface evolution is the latitudinal differential rotation, which effect remains fairly limited over the time span of our observations, and which is at a level similar to the one measured in 2006. We note, however, a systematically larger \chir\ in 2018 compared to 2006 (for all mapping inversions presented here, and in spite of a slightly shorter time span in the 2018 time series), which may suggest a globally higher intrinsic surface variability in 2018. The absence of a clearly correlated evolution between the photosphere and the corona suggests that major reconnexion events can be triggered in the corona without any substantial reorganization of the magnetic field at the surface. It is also possible that very localized surface changes (occuring at spatial scales unresolved through ZDI) are enough to globally alter the stability of the prominence system. This disconnected evolution of the surface and the corona is reminiscent of recent observations of $\epsilon$~Eridani, where a sudden drop in CaII H and K emission was observed within a few days, with no simultaneous changes in the large-scale magnetic field \citep{2021A&A...648A..55P}. 

We estimated the mass of the prominence system by using a method very similar to the one presented by \cite{1996MNRAS.281..626S}, and later used by \cite{2000MNRAS.316..699D} or \cite{2021MNRAS.504.1969Z}. We first calculated the equivalent width (EW) of the emission component of \halpha\ by subtracting from the measured EW a reference EW estimated from a PolarBase spectrum of HD 225261, a quiet star chosen for its low S-index and effective temperature close to the one of \ap\ \citep{2014MNRAS.444.3517M}. By repeating this procedure with $H\beta$ (which DS, showing marginally detected emission signatures consistent with \halpha, is displayed in Fig. \ref{fig:hbeta}), we estimated the Balmer decrement to be equal to $\approx 2.49$, consistent with optically thin material. We make the rough assumption that  the hydrogen clouds are contained within a sphere of radius $l$, taken equal to the radius of the source surface proposed by C20 ($l = 2.5 R_{*}$). The number density of hydrogen atoms is estimated to be $\approx1.1\times10^{16}$~m$^{-3}$ and the total mass stored as prominences is $\approx  4.6 \times10^{17}$~kg ($\approx  2.3 \times10^{-13}M_\sun$). The prominence system mass of V530 Per is, therefore, mostly consisten with the range of masses obtained for other rapidly rotating stars ($10^{-14} - 10^{-17}M_\sun$, \citealt{1989MNRAS.236...57C, 2000MNRAS.316..699D, 2006MNRAS.373.1308D, 2021MNRAS.504.1969Z}), and close to the largest reported value ($\approx  5 \times10^{-14}M_\sun$, \citealt{2000MNRAS.316..699D}). 

While the prominence mass is mostly the same on 17 and 18 Oct, we observe a $\approx 7\%$ increase in the mass measured on 22 Oct. During the last observing night (23 Oct), the prominence mass is back to a value close to the one obtained at the start of the run. This suggests that as much as $3.5 \times 10^{16}$~kg of material has been removed from the system within one day, although it is not possible to determine whether a fraction of this material was sent back to the star, or if it was entirely ejected towards the interstellar medium. Mass estimates from 2006 are consistently below the values reported in 2018, by about 5\%, which may be linked to the globally weaker surface magnetic field measured at this earlier epoch.

\section{Conclusions}
\label{sec:conclusions}
We confirmed the main conclusions of C20 regarding the surface distribution of brightness spots and magnetic regions. A dark spot is again reported at high latitudes, and the magnetic geometry is characterized by a field strength in excess of 1~kG (locally), with a dominant toroidal component. The surface differential rotation is shown to follow a simple solar-like law. Using brightness tracers, its intensity is solar-like, while the measured shear is roughly three times larger using magnetic tracers.   

Two components compose the prominence system accompanying \ap. The first component is confined around the corotation radius and has been observed in 2006 and 2018. We also observed a second, rapidly evolving \halpha\ component, which was much closer to the stellar surface. This second component was absent from our 2006 observations, and was especially intense during the third night of the new run. It was possibly linked to a hot chromospheric point, as its emission source was close to the stellar surface, and not seen in absorption. It was anchored at intermediate to low latitudes, and at the phase of the main extension of the polar spot. An isolated event was also recorded with an extremely short lifetime $<10$~min. These short term events at coronal level do not have any noticeable photospheric counterparts in the spot or magnetic coverage. These observations suggest that prominence systems hovering above the most active stars have a complex structure, spanning a range of spatial scales and lifetimes, and with a possible separation between a near-surface, short lived system coexisting with a dynamically stable ring of material in the vicinity of the corotation radius.   

\begin{acknowledgements} 
We are grateful to the anonymous referee for careful comments that helped to clarify the paper. TC would like to acknowledge financial support from the China Scholarship Council (CSC). JFD acknowledges funding from from the European Research Council (ERC) under the H2020 research \& innovation programme (grant agreement \# 740651 NewWorlds). This research made use of the SIMBAD database operated at CDS, Strasbourg, France, and the NASA's Astrophysics Data System Abstract Service.
\end{acknowledgements}

\bibliographystyle{aa} 
\bibliography{ap149p2_draft.bib} 

\appendix
\section{Observation log}

\begin{table}[h]
\caption{Observation log of V530 Per in 17, 18, 22, 23 Oct 2018). }
          \centering
          \begin{tabular}{@{} cccc @{}}
            \hline
Date (Oct 2018)  & HJD (2458400+) & Phase & Peak S/N\\ 
            \hline
17 & 8.99494 & 0.9655 & 105  \\ 
17 & 9.02392 & 0.0559 & 106  \\ 
17 & 9.05458 & 0.1516 & 105  \\ 
17 & 9.08355 & 0.2420 & 106  \\ 
17 & 9.11393 & 0.3368 & 98   \\ 
17 & 9.14289 & 0.4271 & 97   \\ 
\hline
18 & 9.90980 & 0.8200 &  100  \\ 
18 & 9.93932 & 0.9121 &  100  \\ 
18 & 9.96921 & 0.0053 &   99  \\ 
18 & 9.99874 & 0.0975 &   98  \\ 
18 & 10.02863 & 0.1907 &  99  \\ 
18 & 10.05815 & 0.2828 & 103  \\ 
18 & 10.08808 & 0.3762 & 100  \\ 
18 & 10.11764 & 0.4685 &  98  \\ 
\hline
22 & 13.85090 & 0.1167 & 108  \\ 
22 & 13.88046 & 0.2089 & 102  \\ 
22 & 13.91070 & 0.3033 & 100  \\ 
22 & 13.94026 & 0.3955 &  88  \\ 
22 & 13.97037 & 0.4895 &  94  \\ 
22 & 13.99993 & 0.5817 &  91  \\ 
22 & 14.03189 & 0.6814 &  91  \\ 
22 & 14.06144 & 0.7736 &  95  \\ 
22 & 14.09169 & 0.8680 &  97  \\ 
22 & 14.12124 & 0.9602 &  97  \\ 
\hline
23 & 14.83713 & 0.1939 & 106  \\ 
23 & 14.86665 & 0.2860 & 108  \\ 
23 & 14.89699 & 0.3806 & 110  \\ 
23 & 14.92654 & 0.4728 & 107  \\ 
23 & 14.95677 & 0.5671 & 109  \\ 
23 & 14.98633 & 0.6594 & 110  \\ 
23 & 15.01713 & 0.7555 & 107  \\ 
23 & 15.04668 & 0.8477 & 107  \\ 
23 & 15.07681 & 0.9417 & 108  \\ 
23 & 15.10637 & 0.0339 & 103  \\ 
            \hline
          \end{tabular}  
\begin{tablenotes}
      \small
      \item \textbf{\tt{Note:}}~Every Stokes V spectra consisted of 4 individual unpolarized subexposures with fixed exposure time equal to 600s. From left to right, we list the date, the Julian date, the rotational phases calculated with Eq.\ref{equ:emp}, and the peak S/N.
 \end{tablenotes}
\label{tab:obslog}
\end{table}

\newpage
\section{Example Stokes I LSD profiles}
\begin{figure}[h!]
    \centering
    \includegraphics[width=8cm,trim={2cm 9cm 2cm 9cm},clip]{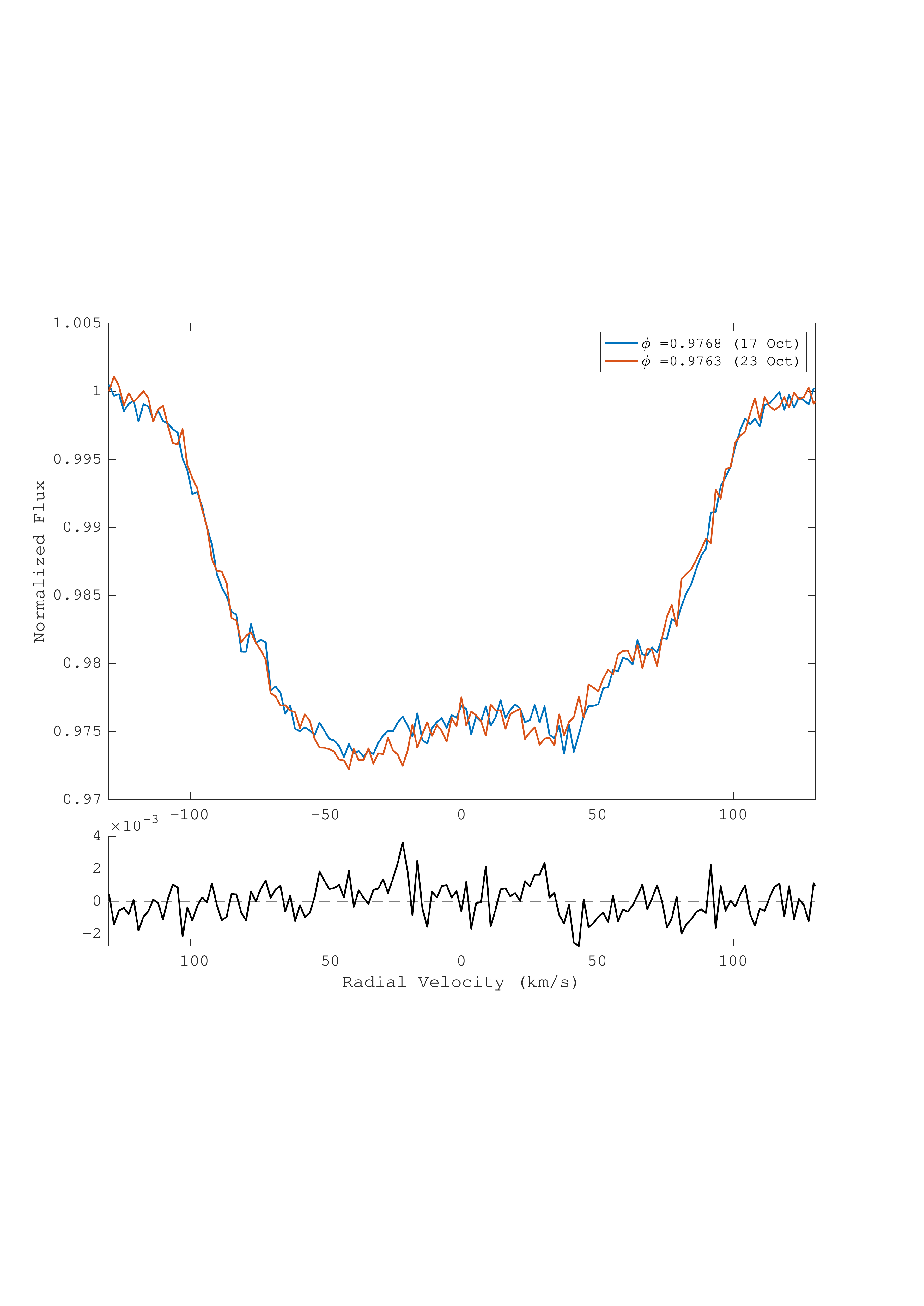}
    \caption{Top panel: example of LSD profiles for 17 Oct. (blue), and 23 Oct. (red) at close rotational phases ($\phi=0.9768$ and $0.9763$). The small differences between the two profiles (e.g., at RV$=-30$~\kms) are consistently observed in other couples and are mainly caused by latitudinal differential rotation. Bottom panel: difference between the two profiles.}
    \label{fig:lsd_stI}
\end{figure}
\clearpage

\section{Dynamic spectrum of \StI~for each night}
\begin{figure}[!h]
    \centering
    \includegraphics[width=1\textwidth,trim={0cm 0cm 0cm 0cm}]{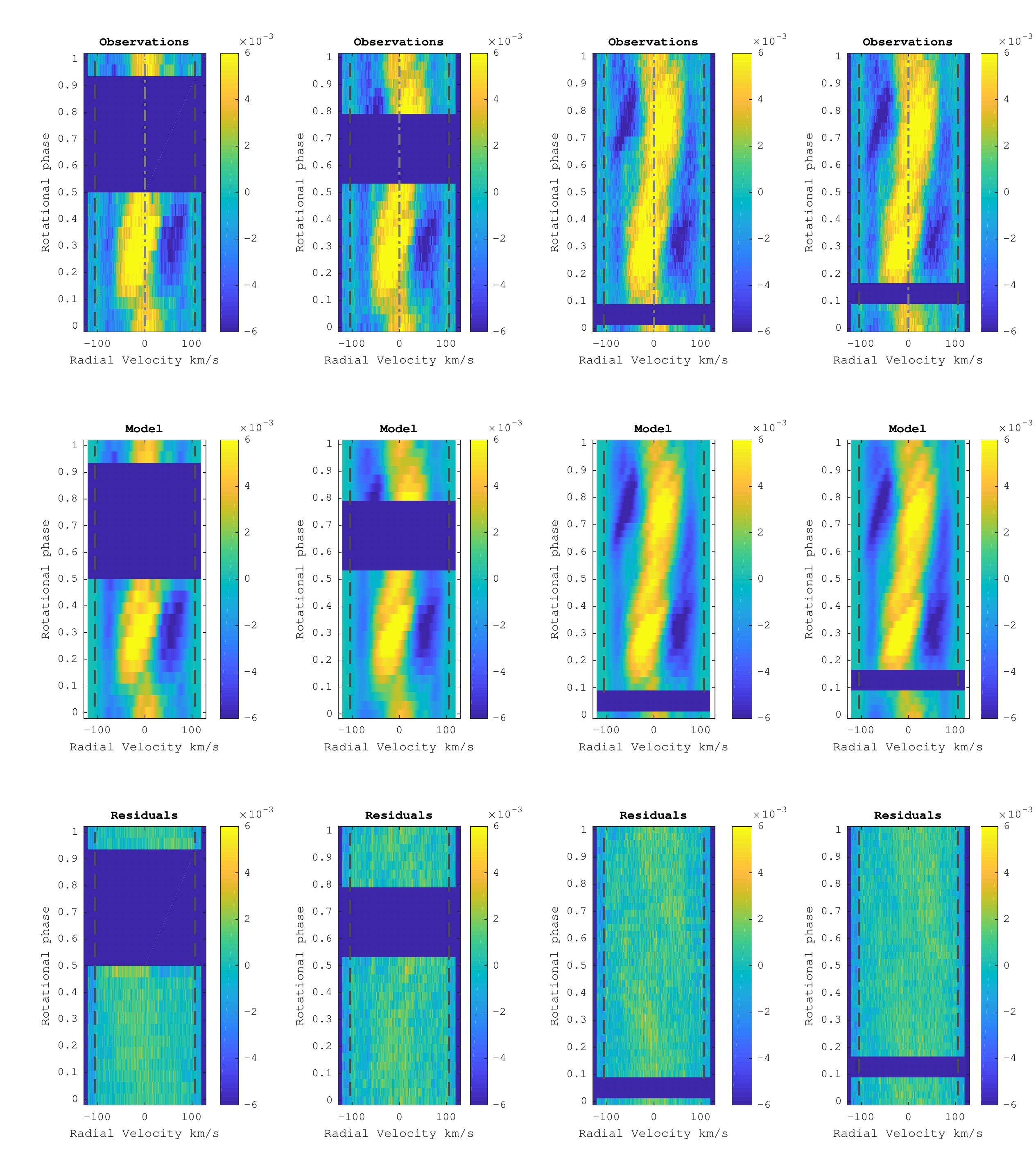}
    \caption{Dynamic spectra for V530 Per. From left to right: 17, 18, 22, 23 Oct. 2018. Upper panels: \StI\ profiles for each night
    Middle panels: ZDI brightness model. Bottom panels: model residuals.}
    \label{fig:DS_stI_4night}
\end{figure}

\clearpage
\section{Dynamic spectrum of \StV}
\begin{figure}[!h]
    \centering
    \includegraphics[width=1.0\textwidth]{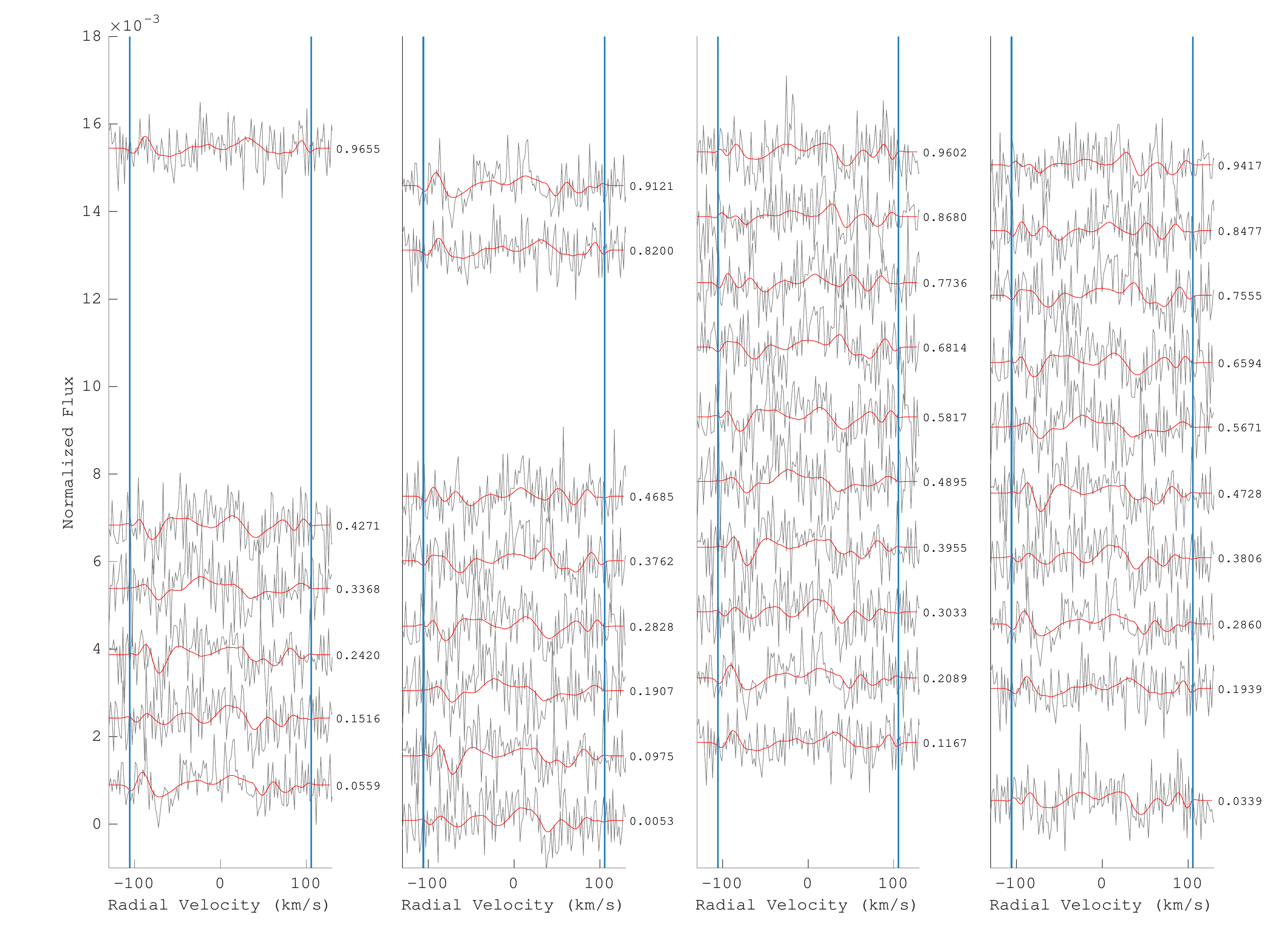}
    \caption{Observed (black) and synthetic (red) Stokes V profiles. The subpanels from left to right show data of the 17, 18, 22, and 23 Oct, respectively. Blue vertical lines mark the $\pm$~\vsini\ limit. Rotational phases are indicated on the right side of each panel, next to the corresponding profile.}
    \label{fig:DS_stV}
\end{figure}{}

\clearpage
\section{Dynamic spectrum of H$\beta$}
\begin{figure}[!h]
    \centering
    \includegraphics[width=1\textwidth,trim={0cm 3cm 1.5cm 1.5cm},clip]{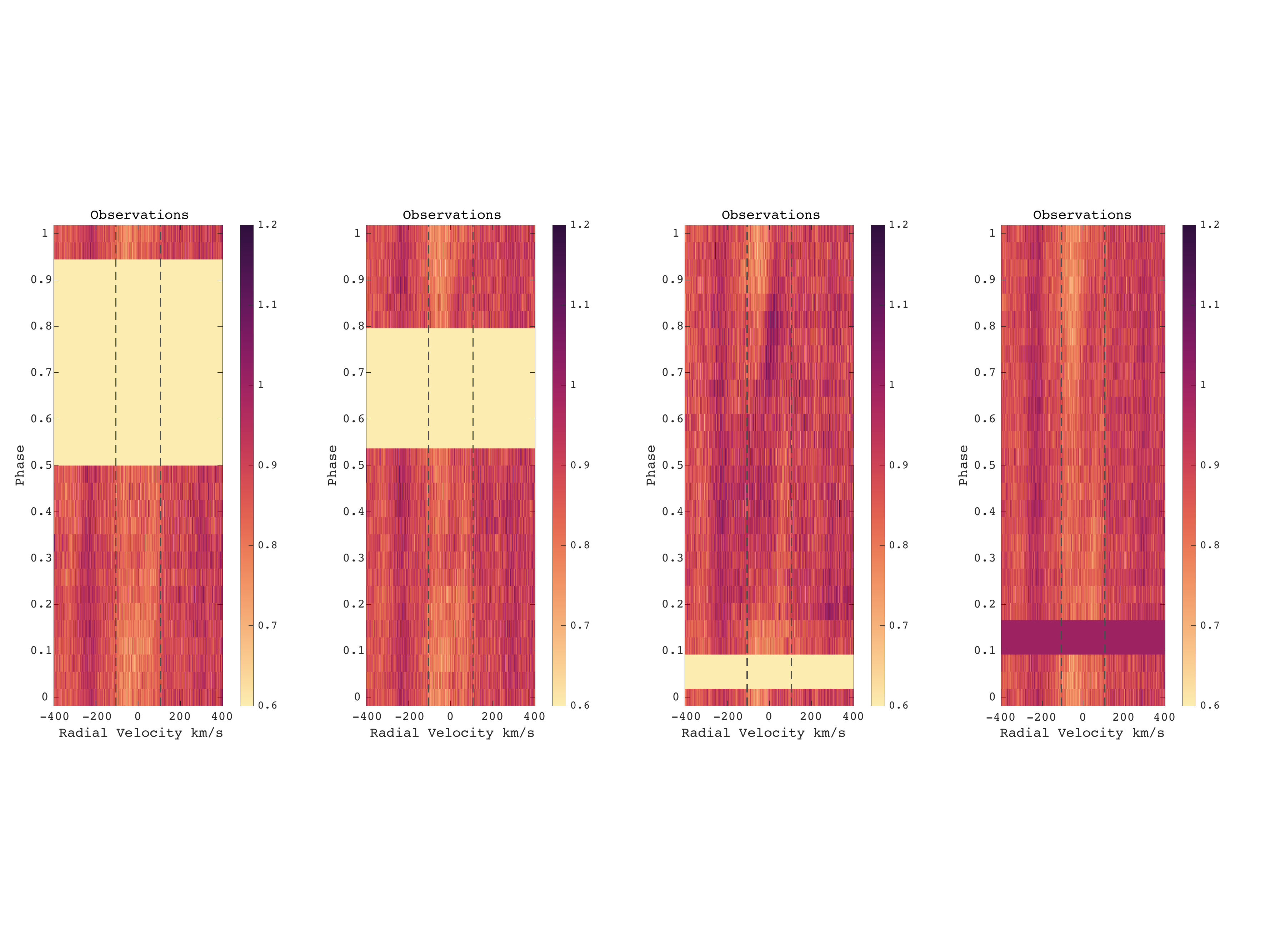}
    \caption{Dynamic spectra of H$\beta$, with the same conventions as in Fig. \ref{fig:DS_halpha}.}
    \label{fig:hbeta}
\end{figure}

\clearpage
\section{Dynamic spectrum of \halpha}
\begin{figure}[!h]
    \centering
    \includegraphics[width=0.9\textwidth,trim={0cm 0cm 0cm 0cm},clip]{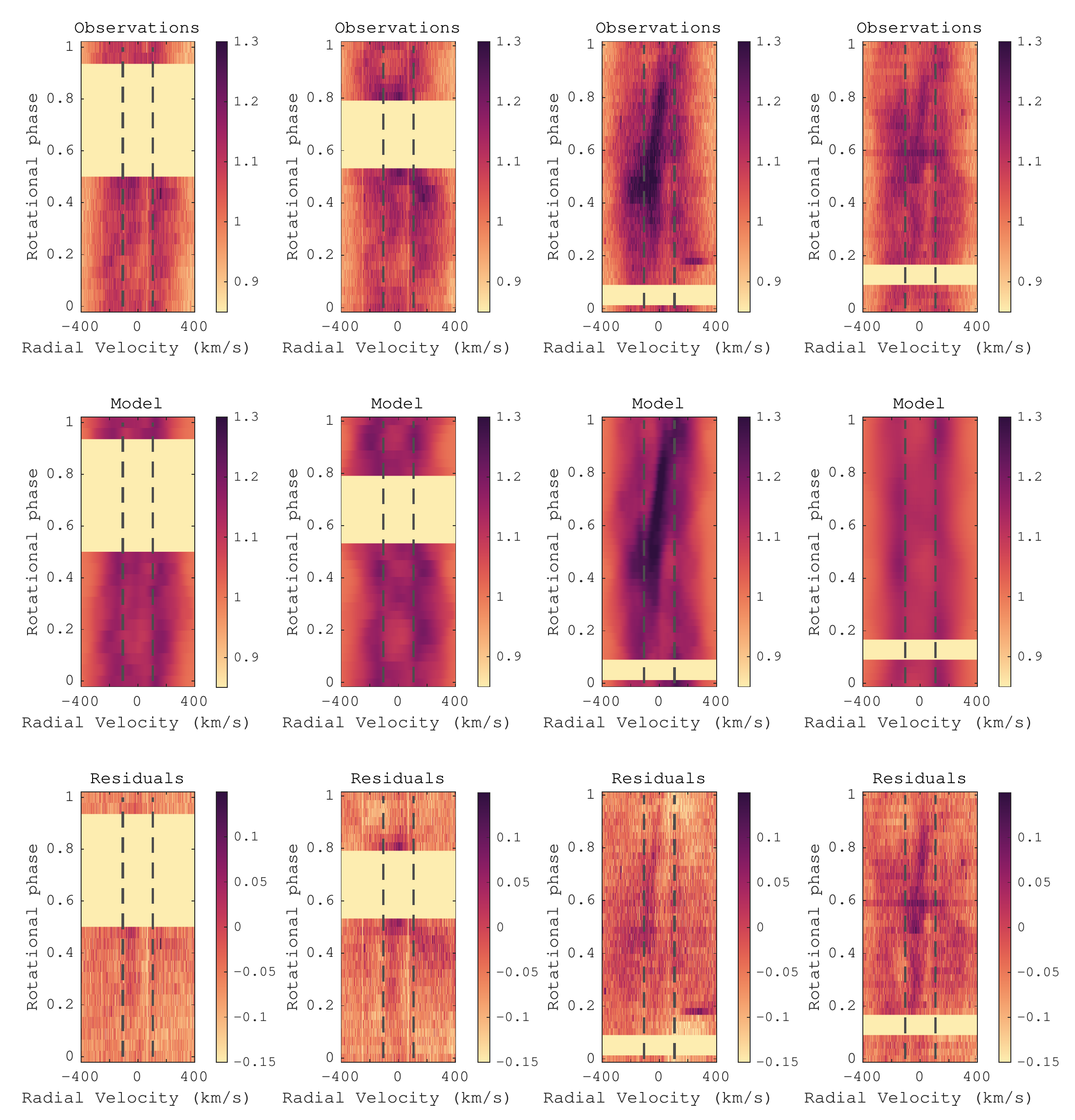}
    \caption{Dynamic spectra for \halpha, including the observations (Top), the tomography models (Middle), the residuals between observation and model (Bottom). From left to right, the figure shows the data of 17, 18, 22, 23 Oct 2018, with color scale according to the normalized flux. Rotational phases are computed according to Eq. \ref{equ:emp}. Vertical dashed lines show the position of $\pm$~\vsini.}
    \label{fig:DS_halpha}
\end{figure}

\clearpage
\section{Polar view of reconstructed brightness and magnetic field maps}
\begin{figure}[!h]
    \centering
    \includegraphics[width=1\textwidth,trim={0cm 0cm 0cm 0cm},clip]{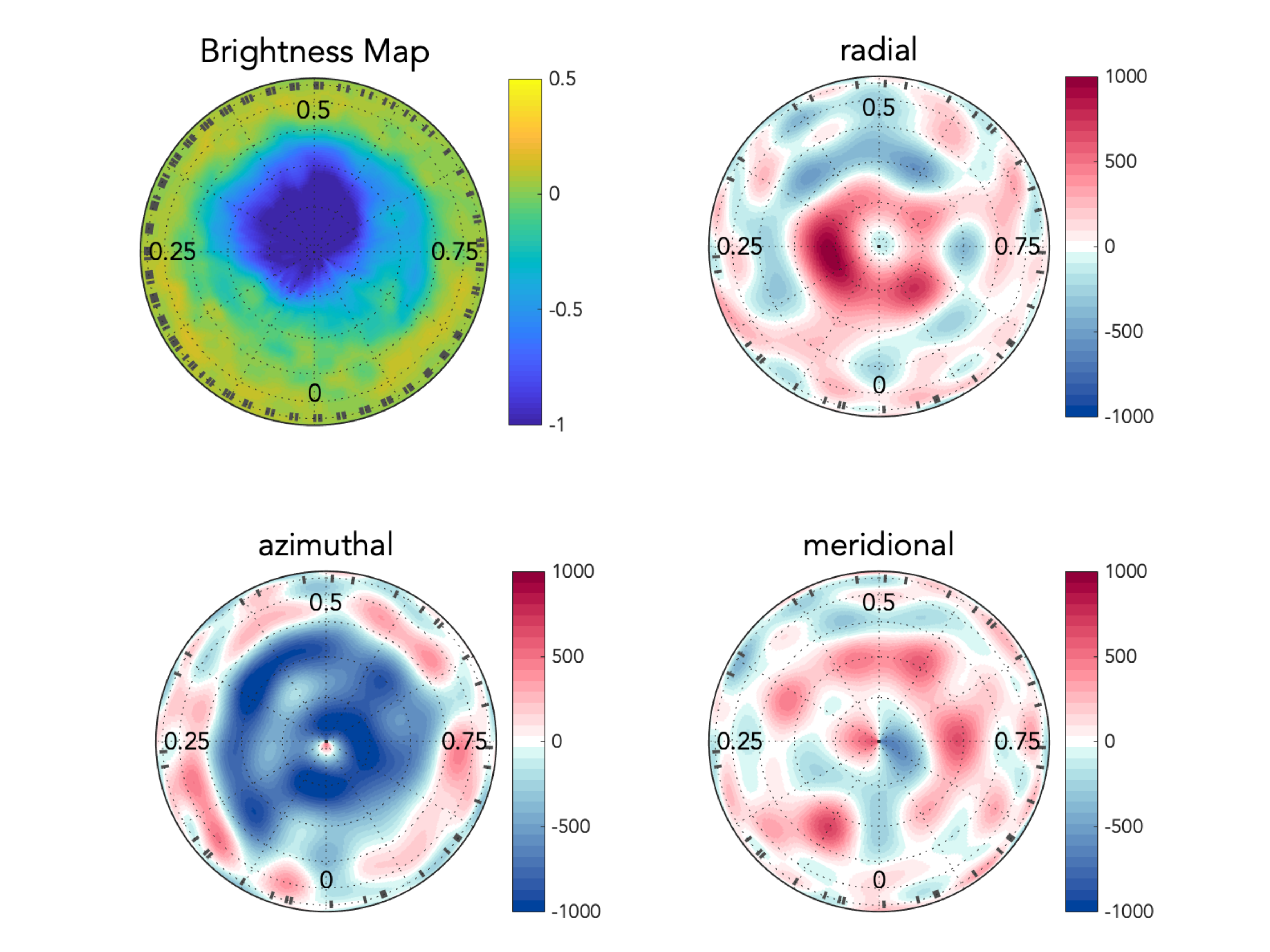}
    \caption{Same as Figure \ref{fig:all_map}, but with a polar view. The color scale of the brightness map (upper left panel) is asymmetric to highlight the low-contrast bright spots.}
    \label{fig:all_map_polar}
\end{figure}{}


\end{document}